\documentclass[a4paper,12pt]{article}

\usepackage{amssymb,amsmath,epsfig}

\newcommand{\A}{{\cal A}}                
\newcommand{\Bra}[1]{\langle\{#1\}|}     
\newcommand{\bra}[1]{\langle#1|}         
\newcommand{\bk}[1]{\langle #1 \rangle}  
\newcommand{\bs}{\boldsymbol}
\newcommand{\C}{\mathbb{C}}              
\newcommand{\cau}{\mathrm{cau}}          
\newcommand{\Dif}[1]{[\mathcal{D} #1 ]}  
\newcommand{\dif}{\mathrm{d}}
\newcommand{\el}{\mathrm{el}}            
\newcommand{\esp}[1]{\mathrm{e}^{#1}}    
\newcommand{\Hc}{\mathcal{H}}
\newcommand{\Ket}[1]{|\{#1\}\rangle}     
\newcommand{\ket}[1]{|#1\rangle}         
\newcommand{\kt}{\boldsymbol{k}}
\renewcommand{\L}{L}             

\newcommand{\ord}[1]{\mathcal{O}\left(#1\right)}
\newcommand{\R}{\mathbb{R}}              
\newcommand{\ret}{\mathrm{ret}}          
\renewcommand{\S}{S}                     
\newcommand{\sign}{\mathrm{sign}}        
\newcommand{\T}{\mathcal{T}}             
\newcommand{\U}{\mathcal{U}}
\newcommand{\ui}{\mathrm{i}}             
\newcommand{\xt}{\boldsymbol{x}}

\title{{\bf Quantum Tunneling and Unitarity Features of an \bf
    $\boldsymbol{S}$-matrix for Gravitational Collapse}}
 
 \author{
   M.~Ciafaloni and D.~Colferai\\[1ex]
   \sl Dipartimento di Fisica, Universit\`a di Firenze\\
   \sl and\\
   \sl INFN, Sezione di Firenze, 50019 Sesto Fiorentino, Italy
 }
 
 \date{}

\begin{document}

 \maketitle
 
 \begin{abstract}
   Starting from the semiclassical reduced-action approach to transplanckian
   scattering by Amati, Veneziano and one of us and from our previous quantum
   extension of that model, we investigate the $S$-matrix expression for
   inelastic processes by extending to this case the tunneling features
   previously found in the region of classical gravitational collapse. The
   resulting model exhibits some non-unitary $S$-matrix eigenvalues for impact
   parameters $b < b_c$, a critical value of the order of the gravitational
   radius $R=2G\sqrt{s}$, thus showing that some (inelastic) unitarity defect is
   generally present, and can be studied quantitatively. We find that $S$-matrix
   unitarity for $b<b_c$ is restored only if the rapidity phase-space parameter
   $y$ is allowed to take values larger than the effective coupling $Gs/\hbar$
   itself. Some features of the resulting unitary model are discussed.
 \end{abstract}
 
 \vskip 10mm
 \begin{minipage}{0.9\textwidth}
 \begin{flushright}
   DFF 451/09/2009~ \\
 \end{flushright}
 \end{minipage}
 \vskip 10mm
 
\section{Introduction}

The ACV eikonal approach to string-gravity at planckian energies~\cite{ACV93}
has been recently investigated in the region of classical gravitational
collapse. A simplified version of it --- the reduced-action model of Amati,
Veneziano and one of us~\cite{ACV07} --- has been extensively studied at
semiclassical level~\cite{ACV07,VW08,OM08}, and has been extended by us (CC) to
a quantum level~\cite{CC08}. The main feature of such a model is the existence
of a critical impact parameter $b=b_c$ of the order of the gravitational radius
$R \equiv 2 G \sqrt{s}$, such that, for $b < b_c$, a classical gravitational
collapse is expected to occur, while the elastic semiclassical $S$-matrix shows
an exponential suppression driven by the effective coupling
$\alpha \equiv G s / \hbar$~\cite{ACV07}. This suppression admits in turn a
tunneling interpretation at quantum level~\cite{CC08}, corresponding to a
partial information recovery, compared to classical information loss.

The purpose of the present paper is to further study the CC quantum model, in
particular its extension to inelastic processes in order to see whether the
tunneling suppression of the elastic channel is possibly compensated by
inelastic production thus recovering $S$-matrix unitarity.

The point above is perhaps the key question that the ACV approach is supposed to
clarify. Indeed, if our $S$-matrix model well represents the original
string-gravity theory, then unitarity is expected irrespective of whether
classical collapse may occur for $b<b_c$. This could be interpreted as full
information recovery at quantum level (compared to classical information loss)
because the suppression of the elastic channel is compensated by the inelastic
ones.

Unfortunately, the situation is not a clearcut one, because of the
approximations involved in the model. On one hand, the reduced-action approach
neglects string and rescattering corrections which --- as argued in~\cite{ACV07}
--- could come in together because of the strong-coupling and, eventually, of
the short distances involved. Furthermore the quantum-extension of~\cite{CC08}
is admittedly incomplete because quantum fluctuations involve only the
transverse-distance dependence of the metric fields, while keeping the classical
shock-wave space-time dependence as frozen. Finally, our extension of the
$S$-matrix to inelastic processes is based on a weak-coupling procedure which
neglects correlations and possible bound states, assumptions which could fail in
a strong-coupling configuration.

Indeed, we find eventually that the model shows a unitarity defect for $b<b_c$,
which is dependent on the rapidity phase-space parameter $y$, in such a way that
unitarity is recovered in the $y\to\infty$ limit only. This result is
interesting because we do have a non-trivial unitary model at large $y$'s and
all $b$'s. But it is puzzling also, because it leaves open the question of
whether, for moderate $y$, one of the simplifying assumptions above went wrong,
or whether instead a unitarity defect is a possible feature of quantum gravity
in the classical collapse region.

In order to introduce the subject properly, we summarize in sec.~\ref{s:raa}
both the semiclassical ACV results for the $S$-matrix and the CC quantum
extension, by emphasizing its tunneling interpretation in the elastic
channel. In sec.~\ref{s:tiea} we derive an improved integral representation of
the CC tunneling amplitude which is applicable for any values of the
$y$-parameter, and we discuss the role of absorption for the various regimes of
the elastic amplitude. We start discussing inelastic processes in
sec.~\ref{s:ipsme}, where we provide two classes of $S$-matrix eigenstates, one
corresponding to a weak-field coherent state which exhibits a unitarity defect
for $b<b_c$, and the other with unitary eigenvalues at all $b$'s, which requires
a suitably chosen strong-field configuration. The ensuing expectations on the
unitarity defect around the elastic channel are compared to the direct
path-integral evaluation of $S^\dagger S$ in sec.~\ref{s:uaf}. We find the
$y$-dependent unitarity defect mentioned previously, that we have quantitatively
evaluated at semiclassical level. We also describe the main features of the
unitary large-$y$ model, by discussing in sec.~\ref{s:disc} possible hints of
further improvements.

\section{The reduced-action approach to gravitational $\bs{S}$-matrix\label{s:raa}}

\subsection{The semiclassical ACV results\label{s:sacvr}}

The simplified ACV approach~\cite{ACV07} to transplanckian scattering is based
on two main points.  Firstly, the gravitational field
$g_{\mu\nu}=\eta_{\mu\nu}+h_{\mu\nu}$ associated to the high-energy scattering
of light particles, reduces to a shock-wave configuration of the form
\begin{subequations}\label{metrica}
  \begin{align}
    h_{--}\big|_{x^+=0} &= (2\pi R)a(\xt)\delta(x^-) \;, \qquad
    h_{++}\big|_{x^-=0} = (2\pi R)\bar{a}(\xt)\delta(x^+)
    \label{hlong} \\
    h_{ij} &= (\pi R)^2\Theta(x^+ x^-)
    \left(\delta_{ij}-\frac{\partial_i\partial_j}{\nabla^2}\right) h(\xt) \;,
    \label{hij}
  \end{align}
\end{subequations}
where $a$, $\bar{a}$ are longitudinal profile functions, and
$h(\xt)\equiv\nabla^2\phi$ is a scalar field describing one emitted-graviton
polarization (the other, related to soft graviton radiation, is negligible in an
axisymmetric configuration).

Secondly, the high-energy dynamics itself is summarized in the $h$-field
emission-current $\Hc(\xt)$ generated by the external sources coupled to the
longitudinal fields $a$ and $\bar{a}$. Such a vertex has been calculated long
ago~\cite{Li82,ABC89} and takes the form
\begin{equation}\label{vertex}
  -\nabla^2 \Hc \equiv \nabla^2 a \nabla^2\bar{a}-\nabla_i\nabla_j a
  \nabla_i\nabla_j\bar{a} \;,
\end{equation}
which is the basis for the gravitational effective action~\cite{Li91,Ki95,Ve93}
from which the shock-wave solution~(\ref{metrica}) emerges~\cite{ACV93}. It is
directly coupled to the field $h$ and, indirectly, to the external sources $s$
and $\bar{s}$ in the reduced 2-dimensional action
\begin{equation}\label{2dimAction}
  \frac{\A}{2\pi Gs} = \int\dif^2\xt\left( a\bar{s}+\bar{a}s-\frac12\nabla a
    \nabla\bar{a}+\frac{(\pi R)^2}{2}\left(-(\nabla^2\phi)^2-2\nabla\phi\cdot
\nabla\Hc\right)\right)
\end{equation}
which is the basic ingredient of the ACV simplified treatment.

\begin{figure}[ht!]
  \centering
  \includegraphics[height=0.15\textwidth]{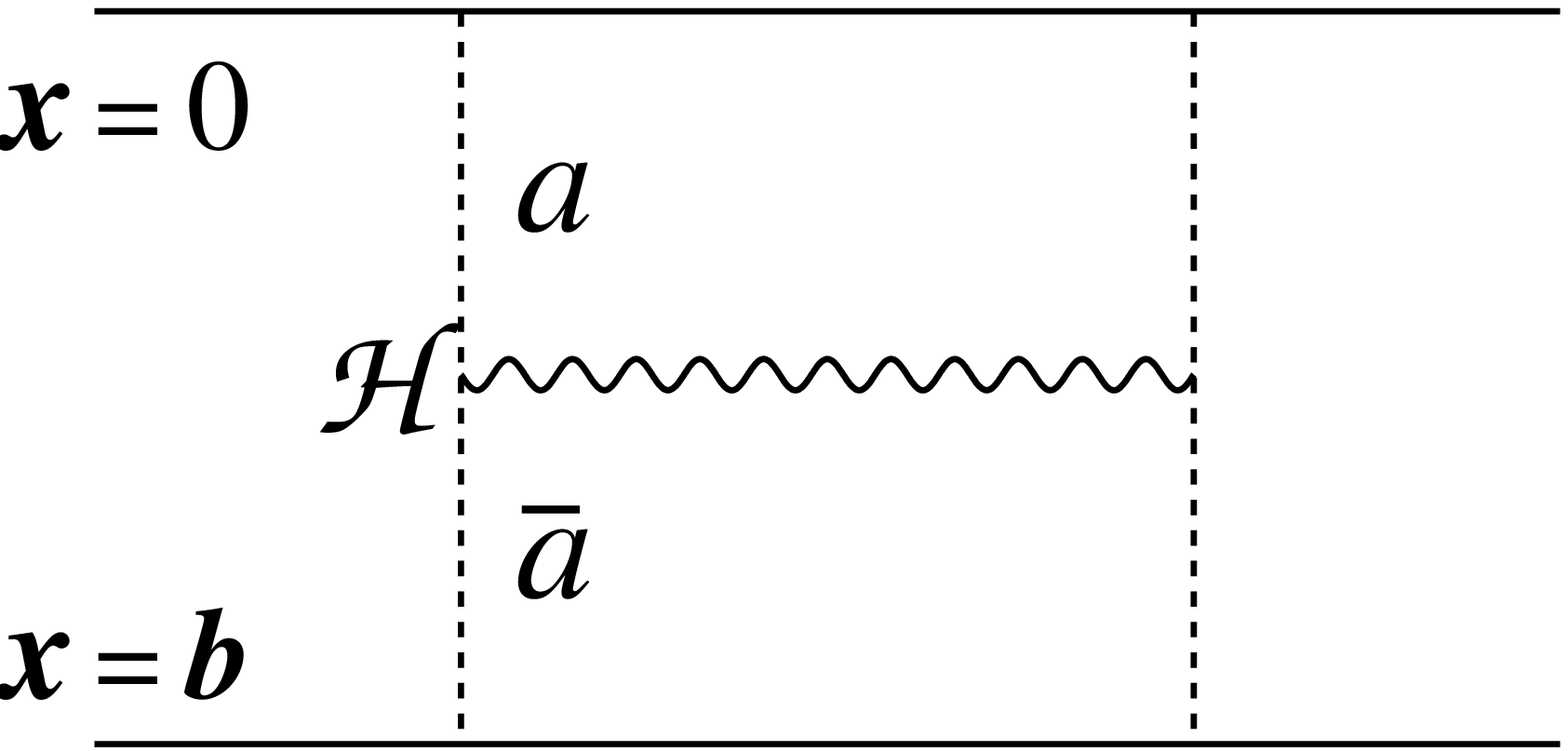}
  \hspace{1em}\raisebox{0.07\textwidth}{+} \hspace{1em}
  \includegraphics[height=0.15\textwidth]{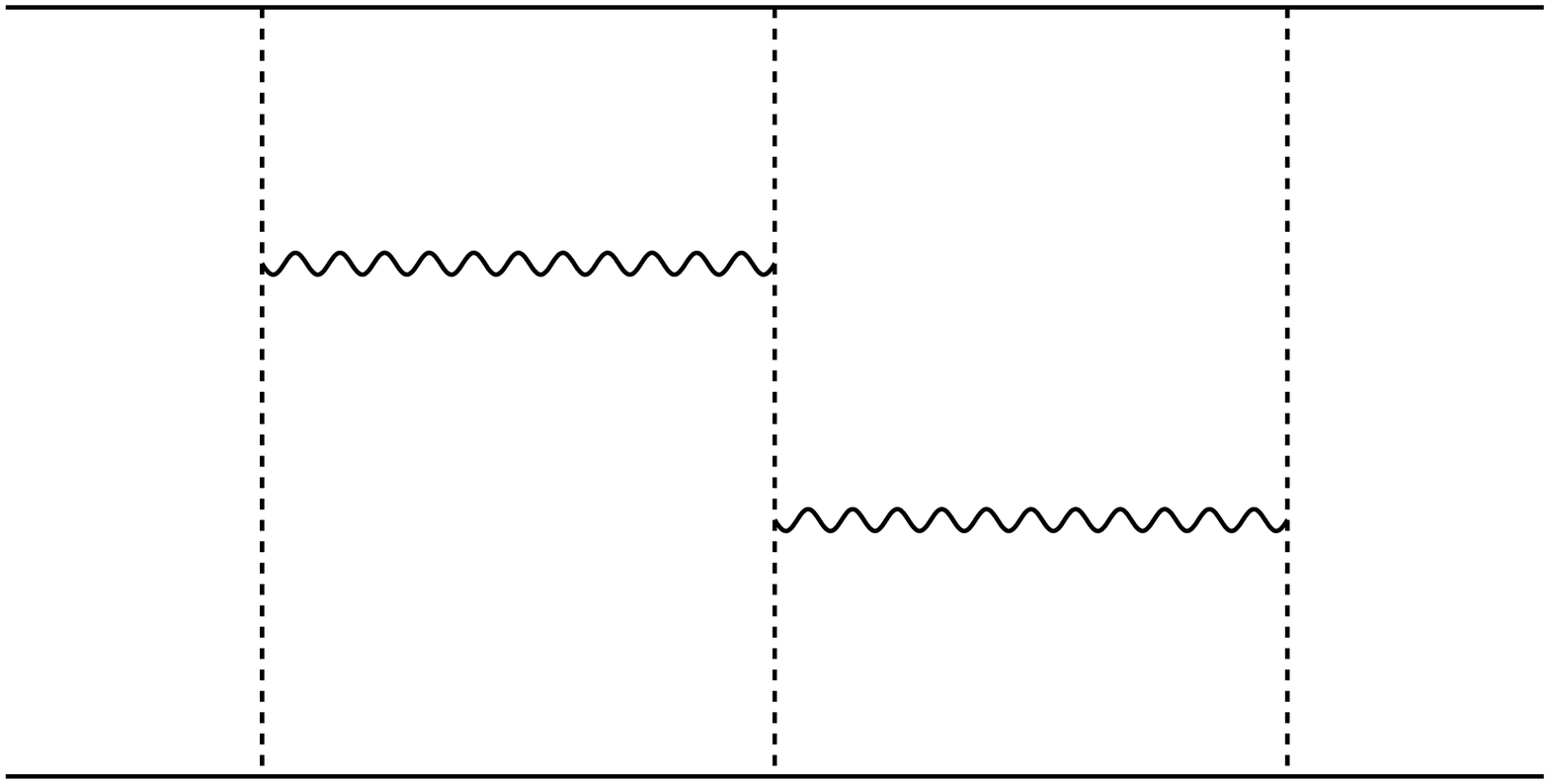}
  \hspace{1em} \raisebox{0.07\textwidth}{+ \dots}
  \caption{\it Diagrammatic series of H and multi-H diagrams.}
  \label{f:multiH}
\end{figure}

The equations of motion (EOM) induced by (\ref{2dimAction}) provide, with proper
boundary conditions, some well-defined effective metric fields $a$ and $h$. The
``on-shell'' action $\A (b,s)$, evaluated on such fields, provides directly the
elastic $S$-matrix
\begin{equation}\label{elSmatrix}
 \S_\el = \exp\left( \frac{\ui}{\hbar} \A(b,s) \right).
\end{equation}
Then, it can be shown~\cite{ACV93,ACV07} that the reduced-action above (where
$R$ plays the role of coupling constant) resums the so-called multi-H diagrams
(fig.~\ref{f:multiH}), contributing a series of corrections
$\sim (R^2/b^2)^n$ to the leading eikonal.

Furthermore, the $S$-matrix (\ref{elSmatrix}) can be extended to inelastic
processes on the basis of the same emitted-graviton field $h(\xt)$.  In the
eikonal formulation the inelastic $S$-matrix is approximately%
\footnote{The coherent state describes uncorrelated emission (apart from
  momentum conservation~\cite{CVprep}). However, the eikonal approach based on
  eq.~(\ref{2dimAction}) also predicts~\cite{ACV93} correlated particle
  emission, which is suppressed by a power of $(Gs/\hbar) Y$ relative to the
  uncorrelated one, and is not considered here.\label{fn:cs}}
described by the coherent state operator
\begin{align}
 &\S = \exp\left( \frac{\ui}{\hbar} \A(b,s) \right)
 \exp\left( \ui 2\pi R \sqrt{\alpha}
 \int\dif^2 \xt \; h(\xt) \Omega(\xt) \right)
 \label{Scsh} \\ \nonumber
 &\Omega(\xt) \equiv \int\frac{\dif^2\kt\,\dif k_3}{2\pi\sqrt{k_0}}
 \left[ a(\kt,k_3) \esp{\ui\kt\cdot\xt} + h. c. \right]
 \equiv A(\xt)+A^\dagger(\xt)\;,\\
 &[ A(\xt), A^\dagger(\xt')] = Y \delta(\xt - \xt')
 \label{Omega}
\end{align}
where the operator $\Omega(\xt)$ incorporates both emission and absorption of
the $h$-fields and $Y$ parameterizes the rapidity phase space which is
effectively allowed for the production of light particles (e.g.~gravitons).

In the following we take the liberty of considering $Y$ as a free, possibly
large parameter which --- for a given value of $\alpha=Gs/\hbar$ --- measures
the longitudinal phase space available. This is a viable attitude at large
impact parameters $b \gg \sqrt{G\hbar}$ because the effective transverse mass of
the light particles is expected to be of order $\hbar/b$, i.e., much smaller
than the Planck mass, thus yielding roughly $Y \gg 1$. On the other hand, we
should notice that dynamical arguments based on energy
conservation~\cite{CVprep} and on absorptive corrections of eikonal type,
consistent with the AGK cutting rules~\cite{AGK73}, tend to suppress the
fragmentation region in a $b$-dependent way, so as to constrain $Y$ to be
$\ord{1}$ for impact parameters in the classical collapse region $b=\ord{R}$.
However, such arguments do not take into account possible dynamical correlations
coming from multi-H diagrams, as mentioned in footnote~\ref{fn:cs}. It is fair
to state that a full dynamical understanding of the $Y$ parameter is not
available yet, and for this reason we shall consider here the full range
$0<Y<\infty$.

In the case of axisymmetric solutions, where $a=a(r^2)$, $\bar{a}=\bar{a}(r^2)$,
$\phi=\phi(r^2)$ it is straightforward to see, by using eq.~(\ref{vertex}), that
$\dot{\Hc}(r^2)\equiv(\dif/\dif r^2)\Hc(r^2)=-2\dot{a}\dot{\bar{a}}$ becomes
proportional to the $a,\bar{a}$ kinetic term. Therefore, the
action~(\ref{2dimAction}) can be rewritten in the more compact one-dimensional
form
\begin{equation}\label{1dimAction}
 \frac{\A}{2\pi^2 Gs}=\int\dif r^2\left( a(r^2)\bar{s}(r^2) + \bar{a}(r^2)s(r^2)
   -2\rho\dot{\bar{a}}\dot{a} - \frac{2}{(2\pi R)^2}(1-\dot\rho)^2\right)
 \;, \qquad \dot{a} \equiv\frac{\dif a}{\dif r^2} \;,
\end{equation}
where we have introduced the auxiliary field $\rho(r^2)$
\begin{equation}\label{rho}
  \rho=r^2\big(1-(2\pi R)^2\dot\phi\big) \;, \qquad
  h=4\dot{(r^2\;\dot{\!\!\!\phi})\!}=\frac1{(\pi R)^2}(1-\dot\rho)
\end{equation}
which incorporates the $\phi$-dependent interaction. The external sources
$s(r^2)$, $\bar{s}(r^2)$ are assumed to be axisymmetric also, and are able to
approximately describe the particle-particle case by setting
$\pi s(r^2)=\delta(r^2)$, $\pi\bar{s}(r^2)=\delta(r^2-b^2)$, where the azimuthal
averaging procedure of ACV is assumed.%
\footnote{The most direct interpretation of this configuration is the scattering
  of a particle off a ring-shaped null matter distribution, which is
  approximately equivalent to the particle-particle case by azimuthal
  averaging~\cite{ACV07}.}

The equations of motion, specialized to the case of particles at impact
parameter $b$ have the form
\begin{align}
  \dot{a} &= -\frac1{2\pi\rho} \;, \qquad
  \dot{\bar{a}} = -\frac1{2\pi\rho}\Theta(r^2-b^2) \;, \label{eoma}\\
  \ddot{\rho} &= \frac1{2\rho^2}\Theta(r^2-b^2) \;, \qquad
  \dot{\rho}^2+\frac1{\rho} = 1 \qquad (r > b) \label{eomrho}
\end{align}
and show a repulsive ``Coulomb'' potential in $\rho$-space, which acts for $r>b$
and plays an important role in the tunneling phenomenon. By replacing the
EOM~(\ref{eoma}) into eq.~(\ref{1dimAction}), the reduced action can be
expressed in terms of the $\rho$ field only, and takes the simple form
\begin{equation}\label{rhoAction}
  \A = -Gs\int\dif r^2
  \left(\frac1{R^2}(1-\dot\rho)^2-\frac1{\rho}\Theta(r^2-b^2)\right)
  \equiv -\int_0^\infty \dif r^2 \; L(\rho,\dot\rho,r^2) \;,
\end{equation}
which is the one we shall consider at quantum level in the following.

Let us now recall the main features of the classical ACV solutions of
eq.~(\ref{eomrho}). First, we set the ACV boundary conditions
$\dot\rho(\infty)=1$ (matching with the perturbative behaviour), and
$\rho(0)=0$, where the latter is required by a proper treatment~\cite{ACV07} of
the $r^2=0$ boundary.%
\footnote{A non-vanishing $\rho(0)$ would correspond to some outgoing flux of
  $\nabla \phi$ and thus to a $\delta$-function singularity at the origin of
  $h$, which is not required by external sources.}
Then, we find the Coulomb-like solution
\begin{align}
  \rho &= R^2\cosh^2\chi(r^2) \;, \qquad \dot\rho = \sqrt{1-\frac{R^2}{\rho}}
   = \tanh\chi(r^2) \equiv t_r \qquad (r^2 \geq b^2) \nonumber \\
  r^2 &= b^2 + R^2(\chi+\sinh\chi\cosh\chi-\chi_b-\sinh\chi_b\cosh\chi_b) \;,
  \label{clSol}
\end{align}
to be joined with the behaviour $\rho=\dot\rho(b^2)r^2$ for $r^2\leq b^2$. With
the short-hand notation $\chi_b\equiv \chi(b^2)$, $t_b\equiv\tanh\chi_b$, the
continuity of $\rho$ and $\dot{\rho}$ at $r^2=b^2$ requires the matching
condition
\begin{equation}\label{crit}
 \rho(b^2) = b^2 \tanh\chi_b = R^2\cosh^2\chi_b \;, \qquad
 \frac{R^2}{b^2} = t_b(1-t_b^2) \;,
\end{equation}
which acquires the meaning of criticality equation.

Indeed, if the impact parameter $b^2$ exceeds a critical value
$b_c^2=(3\sqrt{3}/2)R^2$ at which eq.~(\ref{crit}) is stationary, there are two
real-valued solutions with everywhere regular $\phi$ field, one of them matching
the iterative solution. On the other hand, for $b<b_c$ the ``regular'' solutions
with $\rho(0)=0$ become complex-valued.

The action~(\ref{rhoAction}) evaluated on the equation of motion becomes
\begin{equation}\label{eomAction}
 \frac{\A}{Gs} = \log(4L^2) - \log\frac{1+t_b}{1-t_b} + 1
 -\frac{b^2}{R^2}(1-t_b^2) \;, \qquad (t_b \equiv \tanh\chi_b)
\end{equation}
and provides directly the $b$-dependent eikonal occurring in the elastic
$S$-matrix, while the corresponding $h(r^2)\sim1-\dot\rho$ provides the inelastic
coherent state.

Real-valued solutions for $b<b_c$ exist but are necessarily irregular, i.e.,
$\rho(0) > 0$. Due to the definition of $\rho=r^2[1-(2\pi R)^2\dot\phi]$,
which has the kinematical factor $r^2$, we see that such solutions show a
singularity of the $\dot\phi$ field of type $\dot\phi\simeq-\rho(0)/r^2<0$, so
that one can check~\cite{ACV93} that the metric coefficient $h_{rr}$ must
change sign at some value of $r^2\sim R^2$ and is singular at $r=0$.

A clearcut interpretation of the (unphysical) real-valued solutions with $b<b_c$
and $\rho(0)>0$ is not really available yet. However, we know that in about the
same impact parameter region classical closed trapped surfaces do exist,
as shown in~\cite{EG02,KV02,VW08}. It is therefore tempting to guess that such
field configurations of the ACV approach (which are singular and should have
negligible quantum weight) correspond to classically trapped surfaces. In this
picture, the complex-valued solutions with $\rho(0)=0$ (which are regular, and
should have finite quantum weight) would correspond to the tunneling transition
from the perturbative fields with $\dot{\rho}(\infty)=1$ and positive $\rho$ to
the ``un-trapped'' configuration with $\rho(0)=0$. This suggestion is
incorporated in the quantum level, by defining the $S$-matrix as the
path-integral over $\rho$-field configurations induced by the
action~(\ref{rhoAction}).

\subsection{The quantized CC $\bs{S}$-matrix\label{s:qsm}}

The idea of~\cite{CC08} is to introduce the quantum $S$-matrix as a
path-integral in $\rho$-space of the reduced-action exponential. In this ``sum
over actions'' interpretation the semiclassical limit will automatically agree
with the expression in eq.~(\ref{rhoAction}) above, which is based on the
``on-shell'' action. Furthermore, calculable quantum corrections will be
introduced.

We thus extend the coherent state definition~(\ref{Scsh}) to the quantum level
by introducing it in a path-integral formulation where the
Lagrangian~(\ref{rhoAction}) occurs, as follows
\begin{equation}\label{inSpi}
  \S(b^2,s; \Omega) = \int_{
    \begin{matrix} _{\rho(0)=0} \\ ^{\dot\rho(\infty)=1\;\,} \end{matrix}
  }
  \Dif{\rho(\tau)} \;
 \esp{-\ui\int\dif\tau \; L(\rho,\dot\rho,\tau)} \;
 \esp{\frac{2\ui\sqrt{\alpha}}{\pi R}\int\dif^2\xt\;
 [1-\dot\rho(\tau)]\Omega(\xt)} \;,
\end{equation}
where $\Omega(\xt)$ acts on the multi-graviton Fock space, but is to be
considered as a c-number current with respect to the quantum variables
$\rho,\dot{\rho}$. We also assume the ACV boundary conditions
$\rho(0)=0$, $\dot\rho(\infty)=1$ as discussed above.

In the elastic channel, the $\Omega$-dependent exponential in~(\ref{inSpi}) is
to be replaced by its vacuum expectation value (v.e.v.)
\begin{equation}\label{Sev}
  \exp\left\{-\frac{2Y \alpha}{\pi}\int\dif\tau\; (1-\dot\rho)^2\right\} \;.
\end{equation}
Of course, in this quantum extension, no commitment is made to a particular
classical solution so that the output will presumably contain a weighted
superposition of the various classical paths satisfying the boundary conditions,
that we shall calculate in the following.

Following the above suggestion, we obtain, in the elastic channel,
\begin{equation}\label{Spi}
  \S_\el(b,s) = \int_{
    \begin{matrix} _{\rho(0)=0} \\ ^{\dot\rho(\infty)=1\;\,} \end{matrix}
  }
  \Dif{\rho(\tau)} \;
 \exp\left\{-\frac{\ui}{\hbar}\int\dif\tau \; L_y(\rho,\dot\rho,\tau)\right\} \;.
\end{equation}
where we use the expression~(\ref{rhoAction}) of the reduced action, with the
notations $\tau\equiv r^2$, $y\equiv 2 Y/\pi$ and we introduce the Lagrangian
\begin{equation}\label{lagrangian}
 L_y(\rho,\dot\rho,\tau) = \frac1{4G}\left[(1-\ui y)(1-\dot\rho)^2
 - \frac{R^2}{\rho}\Theta(\tau-b^2)\right] \;, \qquad
 L_{y=0} \equiv L \;,
\end{equation}
with the boundary conditions $\rho(0)=0,\;\dot\rho(\infty)=1$ introduced by ACV
and discussed in sec.~\ref{s:sacvr}.

For generic values of $y$, $L_y$ is complex because of the $(1-\ui y)$ factor in
front of the kinetic term, and is thus able to describe absorptive effects due
to inelastic production. However, in order to deal with a hermitian problem, we
start considering the $y=0$ limit of $\S_\el$ in which $L_y$ is replaced by $L$,
and we shall introduce absorption later on. Although this limit for the elastic
$S$-matrix is somewhat unwarranted --- because absorption turns out to be very
important for unitarity purposes --- we shall see in sec.~\ref{s:ipsme} that the
path-integral~(\ref{Spi}) at $y=0$ acquires the meaning of $S$-matrix eigenvalue
for a class of eigenstates close to the vacuum state.  Therefore, it is anyway
important to discuss it separately.

\subsection{Elastic $\boldsymbol{S}$-matrix as tunneling amplitude\label{s:esm}}

By then setting $y=0$, we shall see that the definition~(\ref{Spi}) given above
is equivalent, by a Legendre transform and use of the Trotter
formula~\cite{RS72}, to quantize the $\tau$-evolution Hamiltonian $H(\tau)$ to
be introduced shortly, and to calculate the evolution operator $\U(0,\infty)$,
thus reducing the $S$-matrix calculation to a known quantum-mechanical problem.
In fact, by eq.~(\ref{lagrangian}), we can introduce the ``conjugate momentum''
\begin{equation}\label{conjMom}
  \Pi \equiv \frac{\partial L}{\partial\dot\rho} = \frac1{2G}(\dot\rho-1)
\end{equation}
and we obtain
\begin{equation}\label{hamiltonian}
  H(\tau) \equiv \Pi\dot\rho - L = \frac1{4G}\left((\dot\rho)^2-1
  +\frac{R^2}{\rho}\Theta(\tau-b^2)\right) \;, \qquad \dot\rho = 1+2G{\Pi}
\end{equation}
from which the classical EOM~(\ref{eomrho}) can be derived. Then, quantizing the
evolution according to eq.~(\ref{Spi}) amounts to assume the canonical
commutation relation
\begin{equation}\label{cr}
  [\rho,\Pi] = \ui\hbar \;, \qquad
  \dot\rho = -2{\ui\hbar}G\frac{\partial}{\partial\rho}
  \equiv -\frac{\ui R^2}{2\alpha}\frac{\partial}{\partial\rho} \;, \qquad
  \alpha\equiv\frac{Gs}{\hbar}
\end{equation}
and to quantize the Hamiltonian~(\ref{hamiltonian}) accordingly:
\begin{equation}\label{Hquant}
 \frac{\hat{H}}{\hbar} = -\frac{R^2}{4\alpha} \frac{\partial^2}{\partial\rho^2}
 + \alpha\left(\frac{\Theta(\tau-b^2)}{\rho}-\frac1{R^2}\right)
 \equiv \frac{H_0}{\hbar}+\frac{\alpha}{\rho}\Theta(\tau-b^2) \;.
\end{equation}
Finally, the path-integral~(\ref{Spi}) for the $S$-matrix without absorption is
related by Trotter's formula to a tunneling amplitude involving the
time-evolution operator $\U(0,\infty)$:
\begin{equation}\label{tunnel}
 \S(b,s)\sim \T(b,\alpha) \equiv \langle\rho=0|\U(0,\infty)|\Pi=0\rangle
 \;, \qquad H_0 |\Pi=0\rangle = 0 \;,
\end{equation}
where the initial (final) state expresses the boundary condition
$\dot\rho(\infty)=1$ ($\rho(0)=0$) and $\U(\tau,\infty)$ is the evolution
operator in the Schr\"odinger picture, calculated with $\tau$-antiordering. The
result~(\ref{tunnel}) expresses the elastic $S$-matrix as a quantum mechanical
amplitude for tunneling from the state $\ket{\Pi=0}$ at $\tau=\infty$ to the
state $\ket{\rho=0}$ at $\tau=0$.

We note that the commutation relation~(\ref{cr}) does not follow from first
principles, but is simply induced by the path-integral definition (\ref{Spi}).
Note also that here we allow fluctuations in transverse space, but we keep
frozen the shock-wave dependence on the longitudinal variables $x^{\pm}$. This
means that our account of quantum fluctuations is admittedly incomplete and
should be considered only as a first step towards the full quantum level. This
step, defined by~(\ref{Spi})-(\ref{tunnel}), has nevertheless the virtue of
reproducing the semiclassical result for $\alpha\to\infty$.

A more detailed expression of the tunneling amplitude~(\ref{tunnel}) can be
derived by introducing the time-dependent wave function
\begin{equation}\label{psiTime}
 \psi(\rho,\tau) \equiv \langle\rho|\U(\tau,\infty)|\Pi=0\rangle
\end{equation}
such that
\begin{equation}\label{tunnel_2}
 \T(b,\alpha) \equiv \langle\rho=0|\U(0,\infty)|\Pi=0\rangle =\psi(0,0) \;.
\end{equation}
Since the Hamiltonian~(\ref{hamiltonian}) is time-dependent, the expression of
the wave function at time $\tau\equiv r^2$ is related to the evolution due to
the Coulomb Hamiltonian $H_c\equiv H_0+Gs/\rho$ by
\begin{align}
 |\psi(\tau)\rangle &= 
 \exp\left(\frac{-\ui H_c \tau}{\hbar}\right) \U_c(0,\infty) |\Pi=0\rangle
 \qquad (\tau\geq b^2) \label{psiIg} \\
 &=  \exp\left(\frac{\ui H_0 (b^2-\tau)}{\hbar}\right)
 \exp\left(\frac{-\ui H_c b^2}{\hbar}\right) \U_c(0,\infty) |\Pi=0\rangle
 \qquad (\tau < b^2) \label{psiIl} \;.
\end{align}
where, according to eq.~(\ref{Hquant}), we have used ``free'' evolution for
$\tau<b^2$.  Therefore, the tunneling amplitude is obtained by setting $\tau=0$
in eq.~(\ref{psiIl}) as follows
\begin{align}
 \T(b,\alpha) &= \langle\rho=0|\psi(0)\rangle = \langle\rho=0|
 \exp\left(\frac{\ui H_0 b^2}{\hbar}\right)
 \exp\left(\frac{-\ui H_c b^2}{\hbar}\right) \U_c(0,\infty) |\Pi=0\rangle
 \nonumber \\
 &= \int\frac{\dif\rho}{(\pi b^2/\ui\alpha)^{1/2}}\;
 \esp{-\ui\alpha(\rho^2/b^2+b^2)}\psi_c(\rho) \;.
 \label{Tval}
\end{align}
This expression is related, by convolution with the free Gaussian propagator, to
the function
\begin{equation}\label{psic}
  \psi_c(\rho) \equiv \langle\rho|\U_c(0,\infty)|\Pi=0\rangle \;,
\end{equation}
which turns out to be a continuum Coulomb wave function with zero energy. In
fact, due to the infinite evolution from the initial condition
$\Pi=0\iff\dot\rho=1$, $\psi_c(\rho)$ is a solution of the stationary Coulomb
problem
\begin{equation}\label{stCoul}
 H_c \psi_c(\rho) = \hbar\left[-\frac{1}{4\alpha}\frac{\dif^2}{\dif\rho^2}
 +\alpha\Big(\frac1\rho-1\Big)\right]\psi_c(\rho) = 0
\end{equation}
with zero energy eigenvalue (where from now on we express $\rho, r^2, b^2$ in
units of $R^2=4G^2s$).  The form of $\psi_c(\rho)$ is better specified by the
Lippman-Schwinger equation
\begin{equation}\label{LSeq}
 \psi_c(\rho) = \esp{2\ui\alpha\rho} + \alpha G_0(0) \;
 \mathrm{pv}\!\!\left( \frac1{\rho} \right) \psi_c(\rho)
 \;, \qquad G_0(E) = [E-H_0+\ui\epsilon]^{-1}
\end{equation}
and thus contains an incident wave with $\dot\rho=1$, plus a reflected wave for
$\rho>0$ and a transmitted wave in the $\rho<0$ region. Note the principal value
determination of $1/\rho$ which is important for hermiticity purposes.

We then conclude that the amplitude~(\ref{tunnel}) is, by eq.~(\ref{Tval}), the
convolution of a gaussian propagator with the Coulomb wave function
$\psi_c(\rho)$, which has a tunneling interpretation with the Coulomb barrier.
In fact, by eq.~(\ref{LSeq}), it contains a transmitted wave in $\rho<0$ (where
the Coulomb potential is attractive) and incident plus reflected waves in
$\rho>0$ (where it is repulsive). Calculating $\psi_c(\rho)$ allows to find an
explicit expression for the tunneling amplitude (sec.~\ref{s:tiea}).

Note that, at $b=0$ we simply have $\T(0,\alpha)=\psi_c(0)$, so that the
tunneling interpretation is direct and recalls the well-known problem of
penetration of the Coulomb barrier in nuclear physics~\cite{LL}. On the other
hand for $b>0$, the convolution with the free propagator changes the problem
considerably, and is the source of the critical impact parameter, as we shall
see below.

\section{Tunneling interpretation and elastic amplitude\label{s:tiea}}

The main purpose of this section is to improve the similar calculation
of~\cite{CC08}, by obtaining an integral representation of the amplitude which
is valid for any values of $b$ and $y$, even the large-$y$ region which is
important for unitarity purposes (see sec.~\ref{s:ipsme}).

We start calculating the tunneling amplitude (\ref{Tval}) without absorption in
terms of the wave function (\ref{psic}). We shall then introduce absorption
according to the definition (\ref{inSpi}), by discussing in particular the
$S$-matrix in the elastic channel.

\subsection{Basic tunneling wave function\label{s:btf}}

The explicit solution of (\ref{stCoul}) is given by a particular confluent
hypergeometric function of $z\equiv -4\ui\alpha\rho$ defined as follows
\begin{align}
  \psi_c &= N_c \, z \, \esp{-z/2} \Phi(1+\ui\alpha, 2,z) \;, \qquad
  z\Phi''+(2-z)\Phi'-(1+\ui\alpha)\Phi=0
  \nonumber \\
  \Phi&\simeq z^{-(1+\ui\alpha)}\big(1+O(1/z)\big) \;,
 \qquad(\ui z\sim\rho\rightarrow-\infty)
 \label{hyper}
\end{align}
where $\Phi$ is defined in terms of its asymptotic power behaviour for
$\rho\rightarrow -\infty$ and the normalization factor $N_c$, to be found below,
is chosen so as to have, asymptotically, a pure-phase incoming wave for
$\rho\simeq L^2\gg 1$, $L^2$ being an IR parameter used to factorize the Coulomb
phase. We shall call this prescription as the ``Coulomb phase'' normalization at
$b=\infty$.

Here we note that the value $c=2$ in $\Phi(1+\ui\alpha, c,z)$ yields a
degenerate case for the differential equation in (\ref{hyper}) in which the
standard solution with the $\rho\to-\infty$ outgoing wave, usually called
$U(1+\ui\alpha, 2, z)$~\cite{AS}, develops a $z=0$ singularity of the form
$A/z+B\log z$. Then, the continuation to $\rho>0$ is determined by requiring the
continuity of wave function and its {\it flux} at $\rho=0$, as is appropriate
for the principal part determination of the ``Coulomb''
singularity~(\ref{LSeq}). The outcome involves therefore an important
contribution at $\rho>0$ of the regular solution $F(1+\ui\alpha, 2, z)$, so that
we obtain
\begin{align}\label{wavef}
 z\esp{-z/2}\Phi &= z \esp{-z/2} \left(U(1+\ui\alpha,2,z)
 +\frac{\ui\pi\Theta(\ui z)}{\Gamma(\ui\alpha)}F(1+\ui\alpha,2,z)\right)
 \\ \nonumber
 &\simeq \esp{(\pi\alpha-z/2)}\cosh(\pi\alpha)z^{-\ui\alpha}
 +\frac{\Gamma(-\ui\alpha)}{\Gamma(\ui\alpha)}\esp{(\pi\alpha+z/2)}
 \sinh(\pi\alpha) (-z)^{\ui\alpha} \qquad (\ui z\to +\infty)
\end{align}
We are finally able to determine the normalization factor $N_c$ and the value of
$\psi_c(0)$, which is finite and non-vanishing, as follows
\begin{equation}\label{psi_0}
 \T(0,\alpha) = \psi_c(0) = \frac{N_c}{\Gamma(1+\ui\alpha)}
 =(4\alpha L^2)^{\ui\alpha}
 \frac{\exp(-\pi\alpha/2)}{\Gamma(1+\ui\alpha)\cosh\pi\alpha}
\end{equation}
a value which is of order $\esp{-\pi\alpha}$, the same order as the wave
transmitted by the barrier.

\subsection{Integral representation of tunneling amplitude at
  $\boldsymbol{b>0}$}
 
For $b>0$, the calculation of $\T$ in (\ref{Tval}) involves a nontrivial
integral, which should be investigated with care. A convenient way to perform
such calculation uses the momentum representation of the Coulomb wave function
$\psi_c$ in which $\dot{\hat{\rho}} \equiv t$ is diagonal. More precisely, from
eqs.~(\ref{conjMom},\ref{cr}) we introduce the representation ($R=1$)
\begin{equation}\label{rhoRep}
 \hat t \equiv \dot\rho = -\frac{\ui}{2\alpha} \frac{\partial}{\partial\rho}
 \qquad \Longleftrightarrow \qquad
 \hat\rho = \frac{\ui}{2\alpha} \frac{\partial}{\partial t} \;.
\end{equation}
The Fourier transform $\tilde\psi_c(t)$ is defined by
\begin{equation}\label{psicTrans}
 \psi_c(\rho) \equiv \int_{-\infty}^{+\infty} \dif t \;
 \esp{\ui 2\alpha \rho t} \, \tilde\psi_c(t) \;.
\end{equation}
From the stationary Hamiltonian~(\ref{stCoul}) in $t$-space
\begin{equation}\label{tHc}
 H_c = \alpha \hbar \left(t^2-1+\frac{2\alpha}{\ui\partial_t}\right) 
\end{equation}
we derive the following differential equation for $\tilde\psi_c(t)$:
\begin{equation}\label{tEq}
 \frac{\partial_t \tilde\psi_c(t)}{\tilde\psi_c(t)} = \frac{2(\ui\alpha-t)}{t^2-1}
 = \frac{\ui\alpha-1}{t-1} - \frac{\ui\alpha+1}{t+1} \;,
\end{equation}
whose general solution is
\begin{equation}\label{tpsic}
 \tilde\psi_c = N(\alpha) \, (t-1)^{\ui\alpha-1} (t+1)^{-\ui\alpha-1} \;.
\end{equation}

\begin{figure}[ht!]
  \centering
  \includegraphics[width=0.95\textwidth]{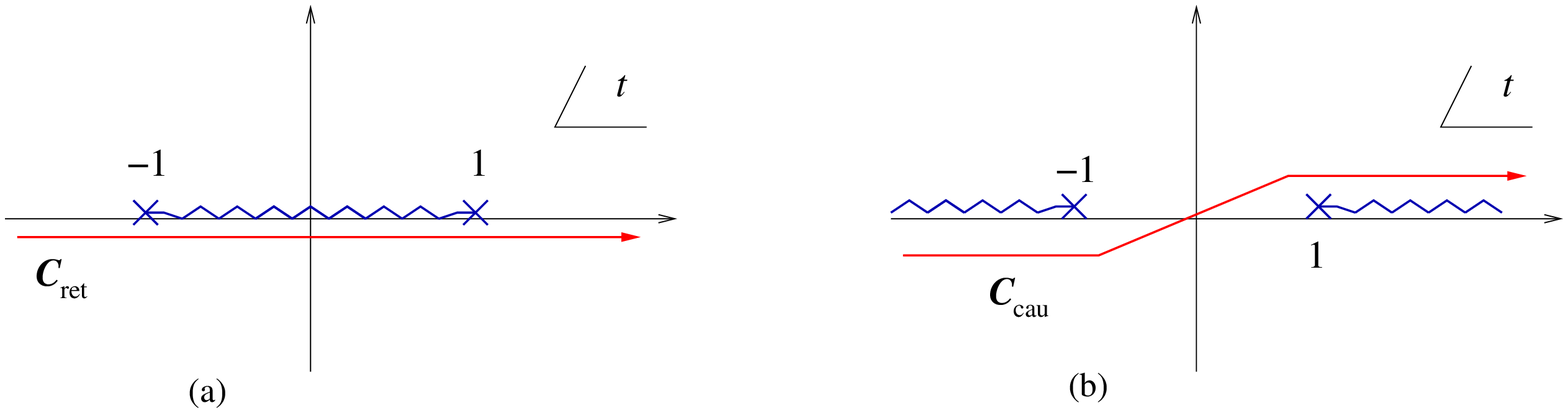}
  \caption{\it Cuts and integration paths for the "retarded" (a) and ``causal''
    (b) solutions of eq.~(\ref{tEq}).}
  \label{f:psiPaths}
\end{figure}

In order to have a meaningful integral in eq.~(\ref{psicTrans}), we need to
shift the singularities of~(\ref{tpsic}) at $t=\pm1$ slightly off the real axis. By
shifting both of them upwards, we obtain an integral representation for the
$F$-part of $\psi_c$:
\begin{align}
 \psi_\ret(\rho) &\equiv \frac{\ui}{\pi} \int_{-\infty}^{+\infty} \dif t \;
 \esp{\ui 2\alpha \rho t}\, (t-1-\ui 0)^{\ui\alpha-1} (t+1-\ui 0)^{-\ui\alpha-1}
 \nonumber \\
 &= \frac{\ui}{\pi} \int_{C_\ret} \dif t \;
 \frac{\esp{\ui 2\alpha \rho t}}{t^2-1} \left(\frac{t-1}{t+1}\right)^{\ui\alpha}
 = \frac{\ui\esp{\pi\alpha}}{\pi} \int_{C_\ret} \dif t \;
 \frac{\esp{\ui 2\alpha \rho t}}{t^2-1} \left(\frac{1-t}{1+t}\right)^{\ui\alpha}
 \nonumber \\
 &= z \esp{-z/2} \Theta(\ui z) F(1+\ui\alpha,2,z) \;,
 \label{psiRet}
\end{align}
where the ``retarded'' subscript, according to standard Green function
notations, indicates that the integration contour lies below the singular points
of the integrand (as shown in fig.~\ref{f:psiPaths}a), yielding a vanishing
result for $\ui z \propto \rho \leq 0$.
The $U$-part of $\psi_c$ can be obtained with a ``causal'' prescription for the
pole shift, as shown in fig.~\ref{f:psiPaths}b:
\begin{align}
 \psi_\cau(\rho) &\equiv  \frac{\ui\esp{\pi\alpha}}{\pi} \int_{-\infty}^{+\infty} \dif t \;
 \esp{\ui 2\alpha \rho t}\, (t-1+\ui 0)^{\ui\alpha-1} (t+1-\ui 0)^{-\ui\alpha-1}
 \nonumber \\
 &= \frac{\ui}{\pi} \int_{-\infty-\ui0}^{+\infty+\ui0} \dif t \;
 \frac{\esp{\ui 2\alpha \rho t}}{t^2-1} \left(\frac{1-t}{1+t}\right)^{\ui\alpha}
 \nonumber \\
 &= z \esp{-z/2} \left[ \frac{\ui}{\pi}\Gamma(\ui\alpha)\sinh(\pi\alpha) U(1+\ui\alpha,2,z)
 + \esp{-\pi\alpha}\Theta(\ui z) F(1+\ui\alpha,2,z) \right] \;.
 \label{psiCau}
\end{align}

The Coulomb wave function~(\ref{wavef}) is now easily obtained as a linear
combination of the retarded and causal solutions:
\begin{align}
 \psi_c &= N_c \frac{\ui\pi}{\Gamma(\ui\alpha)\sinh(\pi\alpha)} \left[
 \cosh(\pi\alpha) \psi_\ret - \psi_\cau \right] \nonumber \\
 &= \frac{(4\ui\alpha L^2)^{\ui\alpha}}{\Gamma(\ui\alpha)\sinh(\pi\alpha)\cosh(\pi\alpha)}
 \left(\int_{C_\cau} - \cosh(\pi\alpha)\esp{\pi\alpha}\int_{C_\ret}\right)
 \frac{\esp{\ui 2\alpha \rho t}}{t^2-1} \left(\frac{1-t}{1+t}\right)^{\ui\alpha}
 \; \dif t \;.
 \label{psicRep1}
\end{align}
A convenient representation of $\psi_c$ with a branch cut at finite $t$ can be
obtained by means of the relation
\begin{equation}\label{cutRel}
 \left(\frac{1-t}{1+t}\right)^{\ui\alpha} = \esp{\sign(\Im t)\,\pi\alpha}
 \left(\frac{t-1}{t+1}\right)^{\ui\alpha} \;,
\end{equation}
and is given by%
\footnote{In order to push the integration paths to the point $t=1$, a
  convergence factor $(t-1)^\epsilon$ must be added to the integrand whenever
  the denominator $t^2-1$ occurs.}
\begin{equation}\label{psicRep2}
 \psi_c = \frac{(4\ui\alpha L^2)^{\ui\alpha}}{\Gamma(\ui\alpha)\cosh(\pi\alpha)}
 \left(-\int_{-\infty-\ui\epsilon}^{1} + \int_{1}^{+\infty} \right)
 \frac{\esp{\ui 2\alpha \rho t}}{t^2-1} \left(\frac{t-1}{t+1}\right)^{\ui\alpha}
 \; \dif t \;.
\end{equation}

It is straightforward at this point to perform the gaussian integration in
eq.~(\ref{Tval})
\begin{equation}\label{gausInt}
 \int\frac{\dif\rho}{(\pi b^2/\ui\alpha)^{1/2}}\;
 \esp{-\ui\alpha(\rho^2/b^2+b^2)} \esp{\ui 2\alpha \rho t}
 = \esp{\ui\alpha b^2(t^2-1)}
\end{equation}
yielding the $b$-dependent tunneling amplitude
\begin{subequations}\label{Tba}
  \begin{align}
    \T(b,\alpha) &= \frac{(4\ui\alpha
      L^2)^{\ui\alpha}}{\Gamma(\ui\alpha)\cosh(\pi\alpha)}
    \left(-\int_{-\infty-\ui\epsilon}^{1} + \int_{1}^{+\infty} \right)
    \frac{\esp{\ui\alpha b^2(t^2-1)}}{t^2-1}
    \left(\frac{t-1}{t+1}\right)^{\ui\alpha} \; \dif t
    \label{Tba1} \\
    &= \frac{(4\ui\alpha L^2)^{\ui\alpha}}{\Gamma(\ui\alpha)\cosh(\pi\alpha)}
    \left(\int_{-\infty-\ui\epsilon}^{1} - \int_{1}^{+\infty} \right) b^2 t
    \left(\frac{t-1}{t+1}\right)^{\ui\alpha} \esp{\ui\alpha b^2(t^2-1)} \; \dif
    t \;,
    \label{Tba2}
  \end{align}
\end{subequations}
where an integration by part has been performed in the last step.%
\footnote{This result is exact, and differs eventually by the integration paths
  from the approximate one in eq.~(4.19) of \cite{CC08}.}

At $b=0$, the transition amplitude can be computed by noting that the integral
$\int_{-\infty-\ui\epsilon}^1+\int_1^{+\infty}$ of the integrand~(\ref{Tba1})
can be closed on the lower half-plane and gives a vanishing result. Therefore
\begin{equation}
 \T(0,\alpha) = \frac{(4\ui\alpha L^2)^{\ui\alpha}}{\Gamma(\ui\alpha)\cosh(\pi\alpha)}
 2\int_{1}^{+\infty}
 \frac{(t-1)^{\ui\alpha-1+0}}{(t+1)^{\ui\alpha+1}} \; \dif t
 = \frac{(4\ui\alpha L^2)^{\ui\alpha}}{\Gamma(\ui\alpha)\cosh(\pi\alpha)} \,
 \frac1{\ui\alpha}
\end{equation}
which correctly reproduces the result in eq.~(\ref{psi_0}).

\begin{figure}[ht!]
  \centering
  \includegraphics[width=0.48\textwidth]{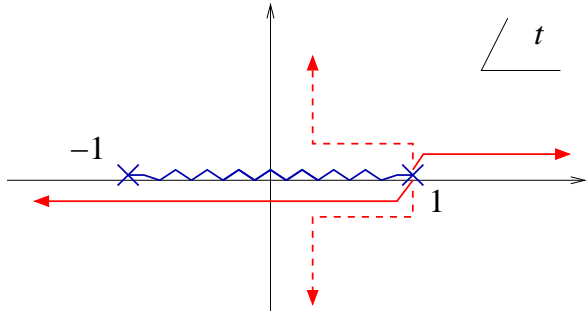}
  \caption{\it Integration paths of eq.~(\ref{Tba}) (solid lines). The corresponding
    deformed paths (dashed lines) are such that the lower one dominates $\T$
    while the contribution stemming from the upper one is strongly suppressed.}
  \label{f:symPaths}
\end{figure}

On the other hand, at $b>0$, the integral $\int_{1}^{+\infty}$ is
exponentially suppressed with respect to $\int_{-\infty-\ui\epsilon}^1$.
This can be shown by bending the paths of the two contributions as shown in
fig.~\ref{f:symPaths} and by noting that the order of magnitude of the integrand
is $\sim\esp{-\pi\alpha}$ above the cut and $\sim\esp{+\pi\alpha}$ below it.

\subsection{Evaluating absorption at quantum level\label{s:iaql}}

In order to take into account multi-graviton emission, we consider the
$S$-matrix in eq.~(\ref{Spi}) with non-vanishing values of the absorption
parameter $y=2Y/\pi$ which effectively takes into account the longitudinal phase
space of gravitons. In the following, we consider $y$ as a free parameter
($0<y<\infty$), independent of $\alpha\equiv Gs/\hbar$, which can vary from
small to large values according to the effective transverse mass of the light
particles being emitted. We note, however, as anticipated in sec.~\ref{s:sacvr},
that the dynamics~(sec.~\ref{s:disc}) will normally introduce correlations, and
the latter can depress or emphasize some regions of rapidity phase space, as it
happens for the case of energy conservation~\cite{CVprep}, thus providing
$\alpha$- and $b$-dependent constraints on the range of possible $y$'s.

For $y\neq0$, the tunneling amplitude with absorption $\T(b,\alpha,y)$ is again
given by eq.~(\ref{tunnel_2}), but in this case the time-dependent wave
function~(\ref{psiTime}) is determined by a non-unitary evolution operator
$\U_y(\tau,\infty)$, due to the fact that the Hamiltonian operator of the
quantum system is no longer hermitian, as it should in order to describe
absorptive effects due to inelastic production.

In fact, the absorption term in eq.~(\ref{inSpi}) adds an imaginary part to the
kinetic term in the Lagrangian~(\ref{lagrangian}) and formally changes the
definition of the Hamiltonian and of the quantization condition in terms of a
new parameter $\tilde\alpha\equiv\alpha(1-\ui y)$:
\begin{equation}\label{Htilde}
 \tilde{H} = \tilde\alpha\left(\dot{\hat{\rho}}\,^2-1\right)
 +\frac{\alpha}{\hat\rho} \Theta(\tau-b^2) \;, \qquad
 [\hat\rho,\dot{\hat{\rho}}]=\frac{\ui\hbar}{2\tilde\alpha} \;, \qquad
 \tilde\alpha \equiv \alpha(1-\ui y) \;.
\end{equation}
A simple way to take into account such changes is to solve the evolution
equation for the wave-function $\langle
t|\psi(\tau)\rangle\equiv\tilde\psi(t;\tau)$ directly in the momentum
representation~(\ref{rhoRep}) in which $\dot{\hat{\rho}}=t$ is diagonal. We
simply obtain
\begin{equation}\label{momSpaceEq}
 \ui\frac{\partial}{\partial\tau}\tilde\psi(t;\tau) = \left[\tilde\alpha(t^2-1)
 +\alpha\Theta(\tau-b^2)\left(\frac{\ui}{2\tilde\alpha}
 \frac{\partial}{\partial t}\right)^{-1}\right] \tilde\psi(t;\tau) \;.
\end{equation}

For $\tau>b^2$, the evolution involves the Coulomb-type Hamiltonian with zero
energy (due to the boundary condition $\dot\rho(\infty)=1$) and we get the
solution
\begin{equation}\label{psit1}
 \tilde\psi(t;\tau) = \left(\frac{t-1}{t+1}\right)^{\ui\alpha} \frac1{t^2-1}
 N(\alpha,y) \;, \qquad (\tau > b^2) \;,
\end{equation}
where the normalization factor $N(\alpha,y)$ will be fixed later on. On the
other hand, for $\tau \leq b^2$ we have just free evolution,
\begin{equation}\label{freeEv}
 \ui\frac{\partial}{\partial \tau} \log\tilde\psi(t;\tau) = \tilde\alpha(t^2-1) \;,
\end{equation}
yielding
\begin{equation}\label{psit2}
 \tilde\psi(t;\tau) = N(\alpha,y) \left(\frac{t-1}{t+1}\right)^{\ui\alpha}
 \frac1{t^2-1} \esp{\ui\alpha(1-\ui y)(1-t^2)(\tau-b^2)} \;, \qquad
 (\tau \leq b^2)
\end{equation}
and therefore
\begin{equation}\label{psiI}
 \psi(\rho,\tau) = N(\alpha,y) \int\dif t\;
 \left(\frac{t-1}{t+1}\right)^{\ui\alpha} \frac1{t^2-1}
 \esp{\ui\alpha(1-\ui y)(1-t^2)(\tau-b^2)}
 \esp{\ui 2\alpha(1-\ui y)\rho t} \;.
\end{equation}
By then setting $\rho=0$ and $\tau=0$, we get the desired result
\begin{equation}\label{psi00}
 \psi(0,0) = N(\alpha,y) \int\dif t\;
 \left(\frac{t-1}{t+1}\right)^{\ui\alpha} \frac1{t^2-1}
 \esp{\ui\alpha b^2(1-\ui y)(t^2-1)} \;,
\end{equation}
which is consistent at $y=0$ with the representation~(\ref{Tba1}), and differs
from it at $y>0$ by the replacement $b^2\to\tilde{b}^2 \equiv b^2(1-\ui y)$.

\begin{figure}[ht!]
  \centering
  \includegraphics[width=0.3\textwidth]{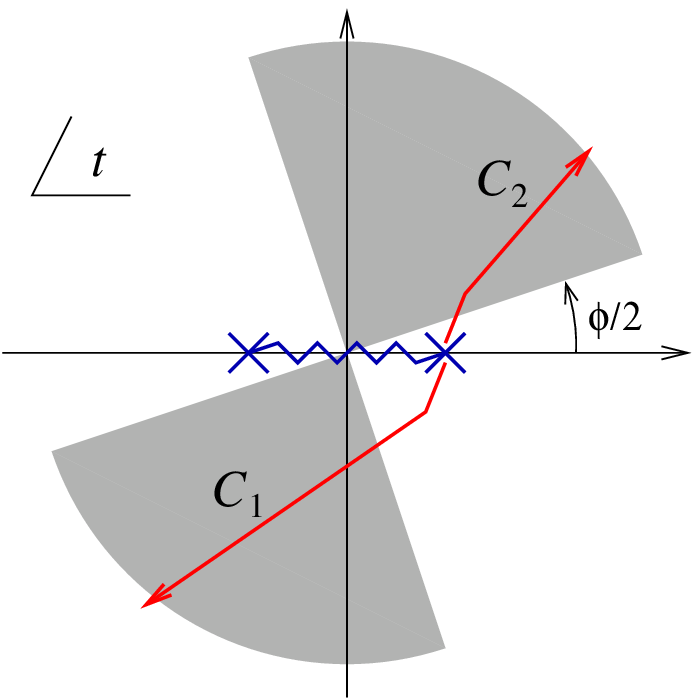}
  \hspace{0.1\textwidth|}
  \includegraphics[width=0.5\textwidth]{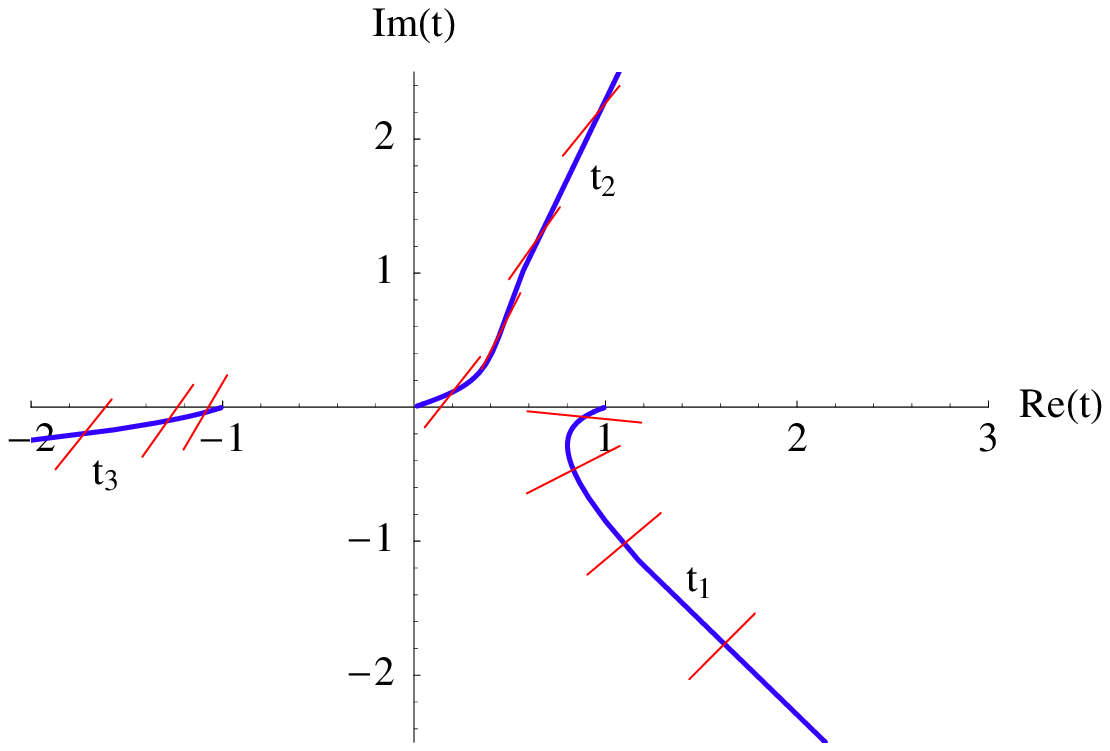}
  \caption{\it (a) Convergence sector and integration paths for the
    $t$-representation of the tunneling amplitude including absorption. (b)
    Position of the saddle points for $y=0.5$ in
    the complex $t$-plane. As $b$ approaches infinity, the three saddle points
    approach the real axis at the points $-1$, $0$ and $1$. The short red lines
    indicate the steepest descent directions for $b=1/4$, $1/2$, $1$ and $2$.}
  \label{f:convSector}
\end{figure}

It remains to determine the proper integration path(s) and the normalization
factor $N$ in eq.~(\ref{psi00}).  In the $y=0$ limit we require $N(\alpha,0)$
and the integration path to agree with eq.~(\ref{Tba1}). By continuity, the
integration path at $y>0$ is obtained by rotating the original one in counter
clock-wise direction, as shown in fig.~\ref{f:convSector}a, in such a way to remain
in the convergence sectors of $\esp{\ui\alpha b^2(1-\ui y)t^2}$, given by
$\phi/2 < \arg(\pm t) < \phi/2+\pi/2$ where $\phi\equiv -\arg(1-\ui y) > 0$.

The normalization factor is fixed by the requirement of unitarity at large $b$,
namely\\
$\lim_{b\to\infty} |\T(b,\alpha,y)| = 1$, and can be determined as
follows. Firstly, one notes that the integrals along $C_1$ and $C_2$
are dominated by saddle points at $t_1$ and $t_2$ respectively, with $t_1 \to 1$
and $t_2 \to 0$ as $b\to\infty$, as shown in fig.~\ref{f:convSector}b.
The saddle point condition is given by $b^2(1-\ui y)t_k(1-t_k^2)=1$ and one has
(cfr.~app.~A of~\cite{CC08})
\begin{align}
 \int_{C_1+C_2}\dif t\;& \left(\frac{t-1}{t+1}\right)^{\ui\alpha} \frac1{t^2-1}
 \esp{\ui\alpha b^2(1-\ui y)(t^2-1)} \nonumber \\
 &\simeq \sum_{k=1}^{2} (-1)^{k-1} t_k
 \left(\frac{t_k-1}{t_k+1}\right)^{\ui\alpha}
 \esp{-\ui\alpha/t_k} \sqrt{\frac{\pi}{\ui \alpha t_k (3 t_k^2 - 1) } }
 \nonumber \\
 & \xrightarrow{b\to\infty}
 \esp{\pi\alpha} \sqrt{\frac{\pi}{2\ui\alpha}}
 \left(4 \esp{} b^2 (1-\ui y)\right)^{-\ui\alpha}
 - \esp{-\pi\alpha} \sqrt{\frac{\ui\pi}{\alpha b^2(1-\ui y)}}
 \esp{-\ui\alpha b^2}\esp{-\alpha b^2 y}
\end{align}
Secondly, one observes that at large $b$ (and even more at large $\alpha$) the
contribution of the saddle point $t_2$ is suppressed with respect to the
contribution from $t_1$, therefore
\begin{equation}\label{TbaySp}
 T(b,\alpha,y) \simeq N(\alpha,y) \, \esp{\pi\alpha} \sqrt{\frac{\pi}{2\ui\alpha}}
 \left(4 \esp{} b^2 (1-\ui y)\right)^{-\ui\alpha}
 \;, \qquad (b\to\infty) \;.
\end{equation}
Finally, from the unitarity requirement, which can be also written as
a ``Coulomb phase'' normalization condition
\begin{equation}\label{coulPhase}
 \lim_{b\to\infty}\frac{\T(b,\alpha,y)}{\T(b,\alpha,0)} = 1 \;,
\end{equation}
we obtain $N(\alpha,y) (1-\ui y)^{-\ui\alpha} = N(\alpha,0)$, and we conclude
that the elastic $S$-matrix (or, the tunneling amplitude including absorption)
is given by
\begin{equation}\label{Tbay}
 \T(b,\alpha,y) = \frac{(4\ui\alpha L^2)^{\ui\alpha}(1-\ui y)^{\ui\alpha}}%
   {\Gamma(\ui\alpha)\cosh(\pi\alpha)}
 \int_{C_1+C_2} \frac{\esp{\ui\alpha b^2(1-\ui y)(t^2-1)}}{t^2-1}
 \left(\frac{t-1}{t+1}\right)^{\ui\alpha} \; \dif t \;.
\end{equation}
Note that the factor $|N(\alpha,y)|=\esp{\alpha\phi}$ is needed to cancel the
extra large-$b$ suppression~(\ref{TbaySp}), and thus enhances the $b=0$
amplitude $\esp{-\alpha(\pi-\phi)}$, which increases to $\esp{-\pi\alpha/2}$ for
$y\to\infty$.

\begin{figure}[ht!]
  \centering
  \includegraphics[angle=-90,width=0.48\textwidth]{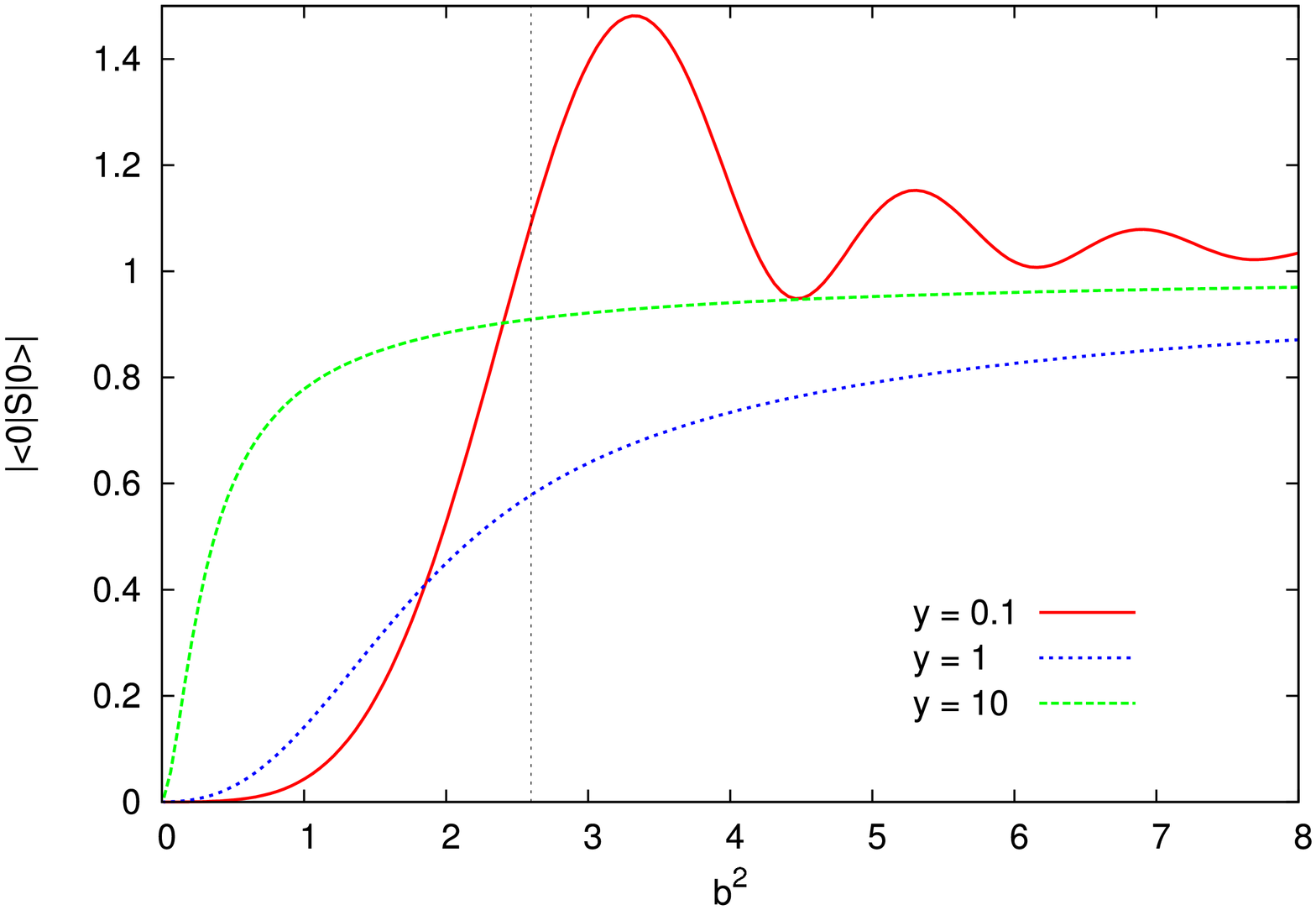}
  \hfill
  \includegraphics[angle=270,width=0.5\textwidth]{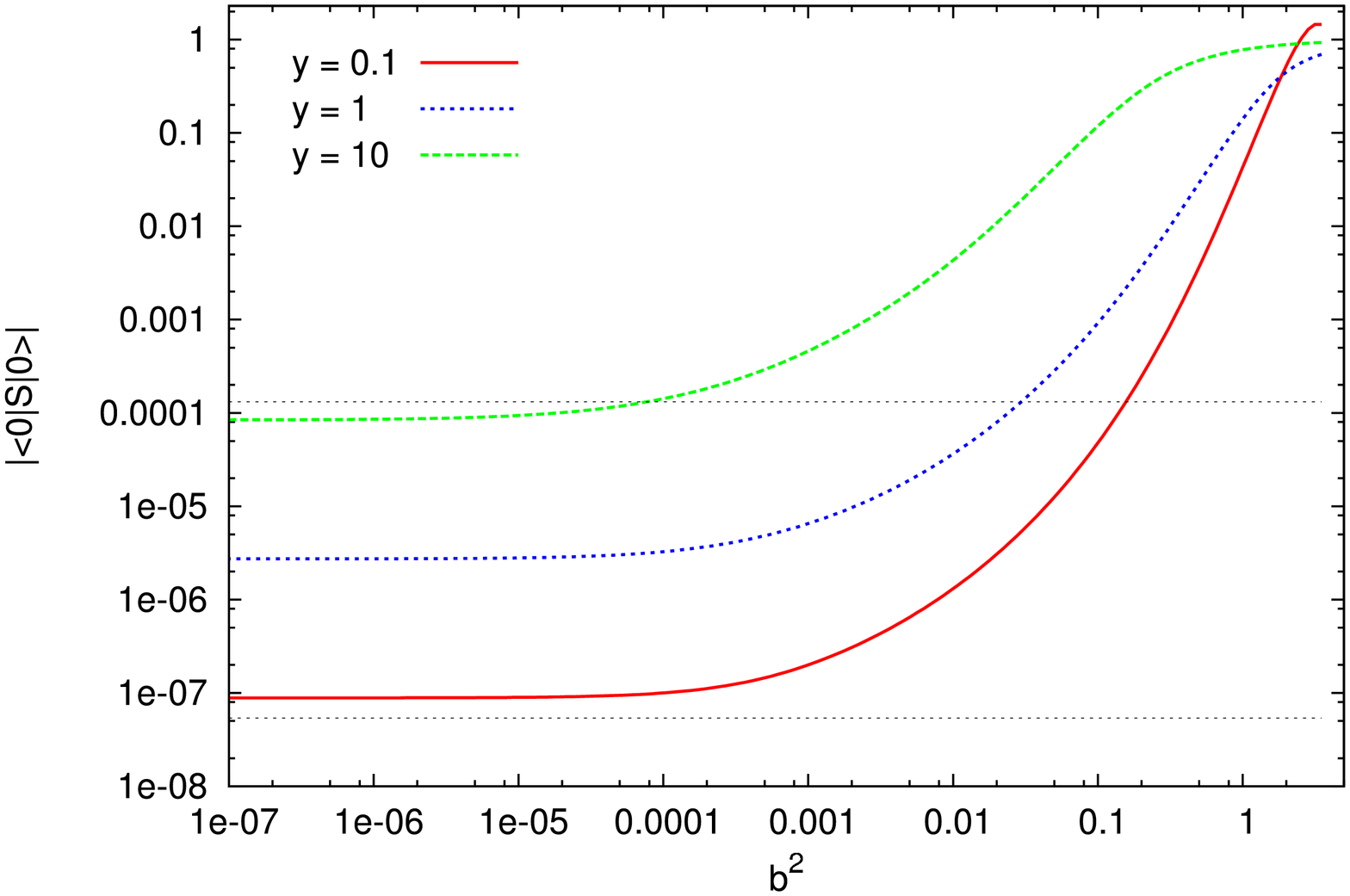}
  \caption{\it Transition amplitude at $\alpha=5$ for three values of the
    absorption parameter y. Large-$b$ behaviour in linear scale (a); small-$b$
    behaviour in logarithmic scale (b). The vertical dashed line in (a) shows
    the critical value $b_c$; the horizontal dashed lines in (b) are the
    boundaries $\sim[\esp{-\pi\alpha},\esp{-\pi\alpha/2}]$ of the $b\to0$ limits
    of the amplitude for $y$ ranging from zero to infinity.}
  \label{f:transAmpl}
\end{figure}

In fig.~\ref{f:transAmpl} we have plotted the dependence on the impact parameter
$b$ of the elastic $S$-matrix, for three values of the inelasticity $y$. For
small $y$'s ($y=0.1$ say) there are some oscillations, due to the interference
of the saddle points $t_1$ and $t_2^{I\!I}$ (on the second $t$-sheet reached
across the $[-1,1]$ cut) for the contour $C_1$. This shows that the elastic
unitarity bound is marginally overcome if $y$ is too small. In all other cases
(with sizeable values of $y$), we observe that the v.e.v.~of
$S$ is below 1 thus satisfying the elastic unitarity bound, and tends to 1 for
large $b$ without oscillations.  This is evidence of only one saddle point
($t_1$) effectively contributing to the integral for sizeable values of $y$ and
$b$.  At larger $y$, fixed $b$, the vacuum-to-vacuum amplitude is less
suppressed than at smaller $y$, and the small-$b$ suppression of the tunneling
amplitude is delayed towards smaller values of $b<b_c$. Roughly, the turning
point is at values of $b$ of order $b_c(1+y^2)^{-1/4}$, thus extending to values
of $b$ smaller than $b_c$ the validity of the perturbative behaviour.
Nevertheless, in the $b\to 0$ limit, the amplitude tends to the
(non-perturbative) constant limit $\esp{-\alpha(\pi-\phi)}$, between
$\esp{-\pi\alpha}$ ($y\to 0$) and $\esp{-\pi\alpha/2}$ ($y\to\infty$).

We thus see the emergence of two absorptive regimes, according to the values of
$y$. In the very small-$y$ regime, quantum interference is important, in
particular for small $b-b_c$ the saddle points $t_1$ and $t_2^{I\!I}$ collide
and interfere by confirming the critical role of $b_c$, but leading to an
analytic $S$-matrix at $b=b_c$, as explained in~\cite{CC08}. On the other hand,
for sizeable to large values of $y$ only one saddle point dominates and the
perturbative and tunneling regimes are hardly distinguishable at $b\simeq b_c$,
the perturbative behaviour with small absorption being extended to smaller
values of $b$. However, we shall see in the following that including inelastic
channels will make things even, by restoring the role of $b=b_c$ for unitarity
purposes, for any values of $y$.

\section{Inelastic processes and $\boldsymbol{S}$-matrix eigenstates\label{s:ipsme}}

So far we have analyzed the $S$-matrix in the elastic channel, deriving in
eq.~(\ref{Tbay}) an explicit expression for the probability amplitude
\begin{equation}\label{vev0}
  \T = \bk{0|S|0}
\end{equation}
which represents, in this simplified model~(\ref{inSpi}) of
transplanckian scattering, the string-string scattering amplitude without
graviton emission (a state represented by the graviton vacuum $\ket{0}$).

We found that starting from the elastic channel (the vacuum state), our quantum
calculation provides absorption for any value of the impact parameter $b$, and
that for $b < b_c$ (critical value) the tunneling absorption persists even if
the graviton-emission phase-space parameter $y$ were set to zero.  This means
that the contribution to the $S$-matrix of quite inelastic states is essential
to possibly recover unitarity.

In this section we investigate the issue of unitarity of our model~(\ref{Scsh})
from various points of view.

\subsection{Eigenstates and eigenvalues of the $\bs{S}$-matrix\label{s:eesm}}

A convenient way to determine whether or not the $S$-matrix is a unitary
operator is to look for its eigenvalues. Due to the particularly simple form of
our $S$-matrix as (superposition of) coherent state operators in the graviton
Fock space, it turns out that the $S$-matrix eigenstates are functional Fourier
transforms of the Fock-space coherent states.  In detail, we define the generic
graviton-coherent-state
\begin{equation}
 \ket{\eta(\tau)} \equiv \esp{-\frac12(\eta^*,\eta)} \exp\left\{
 \int\frac{\dif^2\xt}{\sqrt{\pi}}\;a^\dagger(\xt) \eta(\xt^2) \right\}
 \ket{0}
\end{equation}
where $\eta(\tau)$ is the distribution function of gravitons in the radial
coordinate $\tau\equiv \xt^2$, and we have introduced the scalar product
notation $(\eta,\zeta)\equiv\int_0^\infty\eta(\tau)\zeta(\tau)\;\dif\tau$.
Then, by means of a (normalized) functional integration in $\tau$-space
we introduce the Fourier transform of coherent-states
\begin{equation}\label{Seig}
 \Ket{\omega(\tau)} \equiv \esp{\frac14(\omega,\omega)}
 \int\Dif{\zeta(\tau)}\;\esp{-\ui(\omega,\zeta)}
 \ket{\ui\zeta} \;,
\end{equation}
which are parameterized by the radial function $\omega(\tau)$.  It is
straightforward to prove (app.~\ref{a:esm}) that such states are eigenstates of
the $S$-matrix~(\ref{Scsh})
\begin{align}
  S \Ket{\omega(\tau)} &= \int\Dif{\rho(\tau)} \;
 \esp{-\ui\int \L(\rho,\tau)\;\dif\tau + \ui(\omega,\delta_\rho)}
 \Ket{\omega(\tau)}
 \equiv \esp{\ui\A[\omega;b,\alpha]} \Ket{\omega(\tau)} \;,
 \label{Somega} \\
 \delta_\rho(\tau) &\equiv \sqrt{2\alpha y} \big(1-\dot\rho(\tau)\big) \;,
\end{align}
with eigenvalues $\esp{\ui\A[\omega]}$. Furthermore, the $\omega$-states are
orthonormal in the continuum spectrum and are argued to be complete in the Fock
space~(app.~\ref{a:esm}).

The actual evaluation of the $S$-matrix eigenvalues involves the path-integral
in eq.~(\ref{Somega}), whose action differs from the vacuum one by the
$\omega$-dependent contribution $(\omega,\delta_\rho)$. At the semiclassical
level it is easy to derive the modified equation of motion
\begin{equation}\label{eomW}
  2\ddot\rho -\frac{\Theta(\tau-b^2)}{\rho^2} = -\sqrt{\frac{2y}{\alpha}}
  \dot\omega(\tau)
\end{equation}
in which $\dot\omega$ plays the role of external force, depending on the given
eigenvalue function $\omega(\tau)$.

In the strict $\dot\omega = 0$ limit we are left with the vacuum state equation
characterized by the usual matching condition (in the $y=0$ limit)
\begin{equation}\label{critCond}
  \frac1{b^2} = t_b(1-t_b^2) \;, \qquad \left(t_b = \dot\rho(b^2)\right)
\end{equation}
and by $\rho(b^2) = t_b\, b^2 = \rho_b \equiv 1/(1-t_b^2)$. Real-valued
solutions with $\rho(0)=0$ and $\dot\rho(\infty) = 1$ exist only for
$b\geq b_c$, with $b_c^2 = 3\sqrt{3}/2$. For $b < b_c$ there are complex
solutions, yielding a complex-valued semiclassical eigenvalue and a calculable
absorption, so that $|S(\omega=0;b,\alpha)| < 1$ for $b < b_c$.%
\footnote{We note that the small-$\omega$ solutions with $\Im\A(\omega)>0$ are
  singled out by a stability argument~\cite{ACV07}, so that indeed we can have,
  generally speaking, a unitarity defect and not an overflow.}
This simple observation has the consequence that the $S$-matrix violates
unitarity, to some extent, for values of the impact parameter smaller than the
critical value $b_c$. This means that the $y$-independent $b_c$ separates the
perturbative unitary regime ($b>b_c$) from a regime where a unitarity defect is
possible ($b<b_c$), rather than separating absorptive and tunneling regimes of
the elastic channel, as discussed previously. The actual unitarity violation for
$b<b_c$ is dependent on the relative weight of the small-$\omega$ states in
physical matrix elements and is the subject of the following analysis.

On the other hand, it is essential to note that, if $\dot\omega(\tau)$ is
allowed to take properly chosen (large) values, then real-valued solutions
of~(\ref{eomW}) turn out to {\em exist for all $b$'s}, thus yielding a real
$\A(\omega)$ and a {\em unitary eigenvalue} with $|S(\omega)| = 1$. A large
class ``$R$'' of such solutions is found by setting
\begin{equation}\label{omegaR}
 \omega_R(\tau) = \sqrt{\frac{2\alpha}{y}} \left[ -\frac{\Delta}{1-\Delta}
 (1-\dot\rho_R)\Theta(\tau-b^2) + (B-\dot\rho_R)\Theta(b^2-\tau)\right] \;,
\end{equation}
where $\Delta\in\R$ is arbitrary, $\rho_R$ is the semiclassical solution itself
and $B=(t_b-\Delta)/(1-\Delta)$ by the continuity requirement on $\omega$ and
$\dot\rho_R$ at $\tau=b^2$. By replacing the ansatz~(\ref{omegaR}) in the
equation of motion~(\ref{eomW}) we find in the $\tau > b^2$ region
\begin{equation}\label{eomR}
  2\ddot\rho_R = \frac{1-\Delta}{\rho_R^2} \qquad(\tau > b^2)\;,
\end{equation}
while, for $\tau < b^2$, we can take $\rho(\tau)$ to be any function with
continuous $\rho$ and $\dot\rho$ and finite $\ddot\rho$, satisfying
$\rho(0) = 0$, and matching the Coulomb-like solution in eq.~(\ref{eomR}) at
$\tau = b^2$, i.e., satisfying $\dot\rho(b^2) = t_b$ and
$\rho(b^2) = \rho_b \equiv (1-\Delta)/(1-t_b^2)$. This is an infinite-parameter
set of functions, since the Taylor coefficients $\rho^{(n)}(\bar\tau)$
($0<\bar\tau<b^2$) for $n \geq 3$ are arbitrary.

We see that the effect of the parameter $\Delta$ occurring in the external force
$\dot{\omega}_R$ provided by the eigenstate is to renormalize the Coulomb
coupling in eq.~(\ref{eomR}) by the factor $1-\Delta$, so that it may become
less repulsive for $0\leq\Delta<1$ and even attractive for $\Delta > 1$. The
main point is, though, that eq.~(\ref{eomW}) is identically satisfied by the
ansatz~(\ref{omegaR}) by setting no constraints on $\ddot\rho(\tau)$ in the
$0\leq\tau<b^2$ region, so that the external force allows automatically
real-valued solutions for any value of $b$. Therefore, for any $b$,
eq.~(\ref{omegaR}) yields a family of eigenstates of the $S$-matrix with unitary
eigenvalues depending on an infinite set of parameters: two of them ($\Delta$
and $t_b$) characterize the Coulomb problem in eq.~(\ref{eomR}), and an infinity
of them (the higher-order Taylor coefficients) span the set of functions
$\rho(\tau)$ for $0 < \tau < b^2$.

We stress the point that the very existence of such unitary eigenstates is a
consequence of the quantum structure of the $S$-matrix~(\ref{inSpi}) in which
the field $\rho(\tau)$ is allowed to fluctuate until it reaches the relevant
solution $\rho_R$ of~(\ref{eomW}). The only problem of such states
$\{\omega_R\}$ is that their overlap with the vacuum is suppressed by the factor
\begin{equation}\label{oR0}
  |\Bra{\omega_R} 0\rangle|^2 = \esp{-\frac12(\omega_R,\omega_R)} \;,
\end{equation}
where the exponent is of order $\alpha/y$. Therefore, such states become
important only in the $y \gg \alpha$ limit.

We have thus singled out two families of $S$-matrix eigenstates: the
small-$\omega$ one which exhibits a critical value $b = b_c$, below which no
real-valued semiclassical solutions exist and the tunneling phenomenon occurs
(with non-unitary eigenvalues), and the large-$\omega$ one, in which an
infinite-parameter family of unitary eigenstates exists, characterized by the
eigenvalue functions $\omega_R(\tau)$ in eq.~(\ref{omegaR}). This shows that
unitarity is not an exact property of our quantum model and indicates that
unitarity violations, for any given initial state, are determined by the overlap
profile of such states on the various eigenstates.

\subsubsection{Sum over eigenstates for the elastic channel}

Using the vacuum wave functional
$\Bra{\omega} 0\rangle = \esp{-\frac14(\omega,\omega)}$ it is easy to construct,
by eq.~(\ref{Somega}), the matrix element
\begin{equation}\label{OSomega}
 \bra{0}S\Ket{\omega} = \Bra{\omega}S\ket{0} = \esp{-\frac14(\omega,\omega)}
 \esp{\ui \A(\omega)}
\end{equation}
and then, by summing over the complete set $\Ket{\omega(\tau)}$, the v.e.v.
\begin{align}
 \bk{0|S|0} &= \int\Dif{\omega} \; \bra{0}S\Ket{\omega} \Bra{\omega}0\rangle
 = \int\Dif{\omega} \; \esp{-\frac12(\omega,\omega)}
  \esp{\ui \A(\omega)} \nonumber \\
 &= \int\Dif{\rho} \; \esp{-\ui\int \L(\rho,\tau)\;\dif\tau
  -\frac12 (\delta_\rho,\delta_\rho)} \;,
 \label{OSO}
\end{align}
a result already studied in detail in ref.~\cite{CC08} and in the previous
sections.

We thus remark that the quadratic $\omega$-integration in eq.~(\ref{OSO})
introduces explicitly the absorption parameter $y$ in the vacuum equations, via
the saddle-point value
$\omega_s = \ui\delta_\rho(\tau)=\ui\sqrt{2\alpha y} \big(1-\dot\rho(\tau)\big)$.
We then recover the equation of motion of the elastic channel
\begin{equation}\label{eomE}
 2\ddot\rho(1-\ui y) - \frac{\Theta(\tau-b^2)}{\rho^2} = 0
\end{equation}
whose solutions are complex for any $b$ value, unlike the $\omega = 0$ limit of
eq.~(\ref{eomW}) which admits real-valued solutions for $b > b_c$~\cite{ACV07}.
A consequence of this feature is that for any $b$ value eq.~(\ref{eomE})
predicts the non-vanishing absorption of sec.~\ref{s:iaql}, which, for $b<b_c$,
tends to a finite limit even in the $y = 0$ limit. Therefore, one has to look in
principle at all possible inelastic channels in order to check whether the
absorption of the elastic one can be compensated by the unitarity sum.

\subsubsection{States approximating the unitarity sum\label{s:ssus}}

The simplest approach is to look at the unitarity sum for the $S$-matrix from
the point of view of the squared matrix elements in eq.~(\ref{OSomega}) in order
to identify the states that maximally contribute to the sum. Since the
(quasi)elastic matrix elements are absorbed, and the eigenstates with unitary
eigenvalues are suppressed by the overlap with the vacuum state, the overall
unitarity defect is a balance of the two absorptive effects just mentioned. A
fully quantitative analysis is better done by the method of sec.~\ref{s:uaf}.
Here we look at the contribution of the unitary eigenstates $\Ket{\omega_R}$
only and this will provide a lower bound to the unitarity sum, as follows
\begin{equation}\label{udef}
 \bk{0|S^\dagger S|0} = \int\Dif{\omega}\;|\Bra{\omega}S\ket{0}|^2
 \geq \int\Dif{\omega_R}\;|\Bra{\omega_R}S\ket{0}|^2
 = \int\Dif{\omega_R}\;\esp{-\frac12(\omega_R,\omega_R)} \;,
\end{equation}
where we have used the fact that $|S(\omega_R)|=1$. Thus the suppression
exponent of this lower bound is here provided by the vacuum functional
$(\omega_R,\omega_R)$.

In order to optimize the lower bound above~(\ref{udef}), we look for states that
minimize $(\omega_R,\omega_R)$ in the sample defined by eq.~(\ref{omegaR}). By
imposing stationarity on the infinite set of parameters $\rho^{(n)}(\bar\tau)$
($n\geq3$) we easily find that $\dot\rho(\tau)$ must be a constant for
$\tau<b^2$, and the latter, by continuity, must be $\dot\rho(b^2) = t_b$.
Therefore, we have the matching condition
\begin{equation}\label{match}
 1-\Delta = b^2 t_b (1-t_b^2)
\end{equation}
which corresponds to a Coulomb problem with ``charge'' $(1-\Delta)$. This allows
to replace $\Delta(t_b)$ in the expression
\begin{align}
 \frac12(\omega_R,\omega_R) &= \frac{\alpha}{y} \frac{\Delta^2}{(1-\Delta)^2}
 \left[\int_{b^2}^\infty (1-\dot\rho_R)^2 \;\dif\tau + b^2(1-t_b^2) \right]
 \nonumber \\
 &= \frac{\alpha}{y} \left[ \frac1{b^2} - t_b(1-t_b^2)\right]^2
 \frac{b^2}{t_b^2(1+t_b)} \;.
 \label{oRoR}
\end{align}
For $b \geq b_c$, this expression has a vanishing minimum with $\Delta = 0$,
corresponding to unitarity fulfillment, with a slope parameter varying from
$(1-t_b)\sim 1/(2b^2)$ for $b \gg b_c$, to $t_b = t_c \equiv 1/\sqrt{3}$ for
$b = b_c$ (as usual). Instead, for $b < b_c$, the minimum becomes non-vanishing,
with $t_b = \bar{t}_b$ increasing from $t_c$ to
$\bar{t}_b \sim (b^2)^{-1/3} \to \infty$ for $b$ decreasing from $b_c$ to $0$,
according to the law
\begin{equation}\label{tbmin}
 \frac1{b^2} = \frac{1+3\bar{t}_b}{2+3\bar{t}_b} \bar{t}_b^{\,2} ( 1+\bar{t}_b ) \;.
\end{equation}
Correspondingly, the value of $\Delta$, starting from $\Delta=0$ for $b=b_c$,
increases towards $\Delta=2$ for $b\to0$, so that the Coulomb potential becomes
eventually attractive.  The value of~(\ref{match}) at the minimum becomes
\begin{equation}\label{oRoRmin}
 \frac12(\bar\omega_R,\bar\omega_R) = \frac{4\alpha}{y}
 \frac{(1-3\bar{t}_b^{\,2})^2}{\bar{t}_b^{\,2}(2+3\bar{t}_b)(1+3\bar{t}_b)}
 \xrightarrow{b \to 0} \frac{4\alpha}{y} \;.
\end{equation}
and has the property of vanishing in the $y\to\infty$ limit.

We tentatively conclude that our quantum $S$-matrix is always unitary for
$b>b_c$ and may be unitary for $b<b_c$ also, provided $y/\alpha\to\infty$, the
unitarity sum being approximated by the $\omega_R$'s as given above. This is due
to the fact that $(\omega_R,\omega_R)$ becomes small in that limit, and is
consistent with the vanishing of the ``unitarity action'' $\A_u(y\to\infty)\to0$
that we shall derive in the next section.

\section{The unitarity action and its features\label{s:uaf}}

\subsection{The unitarity action around the vacuum state\label{s:uavs}}

As an alternative method, it is possible to check unitarity directly
by performing the sum over $S$-matrix eigenstates exactly, at fixed field
$\rho(\tau)$. Since the integration over $\omega$ is quadratic, the unitarity
sum becomes
\begin{align}
 \bk{0|S^\dagger S|0} &= \int\Dif{\omega}\; |\Bra{\omega}S\ket{0}|^2
 = \int\Dif{\omega}\; \esp{-\frac12(\omega,\omega)} \esp{-2\Im\A(\omega)}
 \nonumber \\
 &= \int\Dif{\rho}\Dif{\tilde\rho}\;
  \esp{\ui\int[\L(\rho)-\L(\tilde\rho)]\;\dif\tau
  -\frac12(\delta_\rho-\delta_{\tilde\rho},\delta_\rho-\delta_{\tilde\rho})}
  \equiv \int\Dif{\rho}\Dif{\tilde\rho}\; \esp{\ui\A_u} \;,
 \label{OSSO}
\end{align}
where we have performed the $\omega$-integration around the saddle point
$\omega_s=\ui(\delta_\rho-\delta_{\tilde\rho})=\ui\sqrt{2\alpha y}(\dot{\tilde\rho}-\dot\rho)$,
by introducing the path-integral representation of $S(\omega)$. It is then
straightforward to derive the semiclassical equations
\begin{equation}\label{eomU}
  \begin{cases}
    2\ddot\rho - 2\ui y(\ddot\rho - \ddot{\tilde\rho}) &=
    \displaystyle{\frac{\Theta(\tau-b^2)}{\rho^2}} \\[2ex]
    2\ddot{\tilde\rho} + 2\ui y(\ddot{\tilde\rho} - \ddot\rho) &=
    \displaystyle{\frac{\Theta(\tau-b^2)}{\tilde{\rho}^2}}
  \end{cases}
\end{equation}
which govern the unitarity action
\begin{equation}\label{d:Au}
 \A_u \equiv - \int\L(\rho)-\L(\tilde\rho)
  +\ui\alpha y (\dot{\tilde\rho}-\dot\rho)^2 \;\dif\tau \;.
\end{equation}

From eq.~(\ref{eomU}) we see that, for $b > b_c$, real-valued solutions with
$\tilde\rho(\tau)=\rho(\tau)$ exist --- both equations reducing to the elastic
one~(\ref{eomrho}) --- for which the on-shell unitarity action vanishes, thus
implying a unitary $S$-matrix, since, at semiclassical level,
\begin{equation}\label{0SS0sc}
  \bk{0|S^\dagger S|0}_{\mathrm{semicl}} = \esp{\ui\A_u} \;.
\end{equation}
On the other hand, for $b < b_c$, the solutions are necessarily complex and
eq.~(\ref{eomU}) can be satisfied by setting $\tilde\rho = \rho^*$, thus
yielding the equation
\begin{equation}\label{eomI}
 2\ddot\rho + 4 y \Im\ddot\rho = \frac{\Theta(\tau-b^2)}{\rho^2} \;,
\end{equation}
which is equivalent to a coupled set of equations for $\rho_1\equiv\Re\rho$ and
$\rho_2\equiv\Im\rho$. Note that, unlike the elastic channel case, the
equations~(\ref{eomI}) do not have an analytic structure in $\rho$; therefore
they are to be solved as a coupled set of equations having the form
\begin{equation}\label{eom12}
  \begin{cases}
    2\ddot\rho_1 + 4 y \ddot\rho_2 &= \displaystyle{\Re\frac1{\rho^2}
      \Theta(\tau-b^2)} \\[2ex]
    2\ddot\rho_2 &= \displaystyle{\Im\frac1{\rho^2}
      \Theta(\tau-b^2)}
  \end{cases}
\end{equation}
under the boundary conditions
\begin{equation}\label{bc}
 \rho_1(0) = \rho_2(0) = \dot\rho_1(\infty)-1 = \dot\rho_2(\infty) = 0 \;.
\end{equation}

We note that the unitarity action~(\ref{d:Au}) entering the v.e.v.~in
eq.~(\ref{0SS0sc}) can be decomposed into two pieces:
\begin{equation}\label{scomp}
  i\A_u = 2\int\Im L(\rho) \;\dif\tau + 4\alpha y \int h_2^2 \;\dif\tau \;,
  \qquad(h_2 = \Im\dot\rho) \;.
\end{equation}
The first piece is related to the contribution of the vacuum channel ($n=0$) to
the unitarity sum
\begin{equation}\label{unitSum}
 \bk{0|S^\dagger S|0} = \sum_n \bk{0|\S^\dagger|n}\bk{n|\S|0} \;,
\end{equation}
since, by eqs.~(\ref{elSmatrix},\ref{rhoAction}),
\begin{equation}\label{vacuumCont}
 \esp{2\int\Im L(\rho_c)} \simeq |\bk{0|S|0}|^2 = \bk{0|S^\dagger|0}\bk{0|S|0}
\end{equation}
where $\rho_c$ is the Coulomb-like solution~(\ref{clSol}).  The second piece
$\propto h_2^2$ can then be roughly interpreted as the contribution to the
unitarity sum of the inelastic states, and it will be computed in
sec.~\ref{s:nr}.

Some simplification in the discussion of~(\ref{eom12}) is obtained because of
the existence of a constant of motion of energy type. By multiplying the first
equation by $\dot\rho_2$ and the second one by $\dot\rho_1$ and by summing we
easily prove the relation (valid for $\tau \geq b^2$)
\begin{equation}\label{constMot}
 \Im\left((\dot\rho)^2+\frac1{\rho}\right) + 2y\left(\Im\dot\rho\right)^2 =
 2\dot\rho_1\dot\rho_2 - \frac{\rho_2}{|\rho|^2} + 2y(\dot\rho_2)^2 = 0 \;,
\end{equation}
which roughly corresponds to the imaginary part of the single-channel ``energy''
$(\dot\rho)^2+1/\rho$ (in the $y = 0$ limit).

No additional constant of motion seems to be present, the system appearing to be
of dissipative type and thus not integrable analytically. We quote a general
expression for the on-shell unitarity action $\A_u$, derived in app.~\ref{a:ua}:
\begin{equation}\label{Au}
 \ui\A_u(y) = 2\alpha\left( 2\rho_2(\infty) + 3 \Im\frac1{t_b}\right) \;.
\end{equation}
Here $t_b = \dot\rho(b^2)$ and $\rho_2(\infty)$ characterize the given solution,
but do not appear to be related in closed form, so that no matching condition
emerges analytically. Nevertheless, one can argue that $\ui\A_u(y) \leq 0$ with
positive $y$-derivative and that $\lim_{y\to\infty} \A_u(y) = 0$. Indeed, on the
basis of the equations of motion one can show (app.~\ref{s:yinf}) that
\begin{equation}\label{derAu}
 \ui\frac{\dif\A_u(y)}{\dif y} = 4\alpha\int \rho_2^2(\tau)\;\dif\tau > 0
\end{equation}
and that, for large $y$, $y\rho_2(\tau;y)$ reaches a finite limit
$R_2(\tau)$. As a consequence, in eq.~(\ref{Au}) both $\rho_2(\infty)$ and
$t_2\equiv\Im t_b$ are of order $1/y$. It follows that
$|\A_u| = \ord{\alpha/y}$, and thus vanishes in the $y\to\infty$ limit.

\subsection{Numerical results\label{s:nr}}

\begin{figure}[hb!]
  \centering
  \includegraphics[angle=270,width=0.8\textwidth]{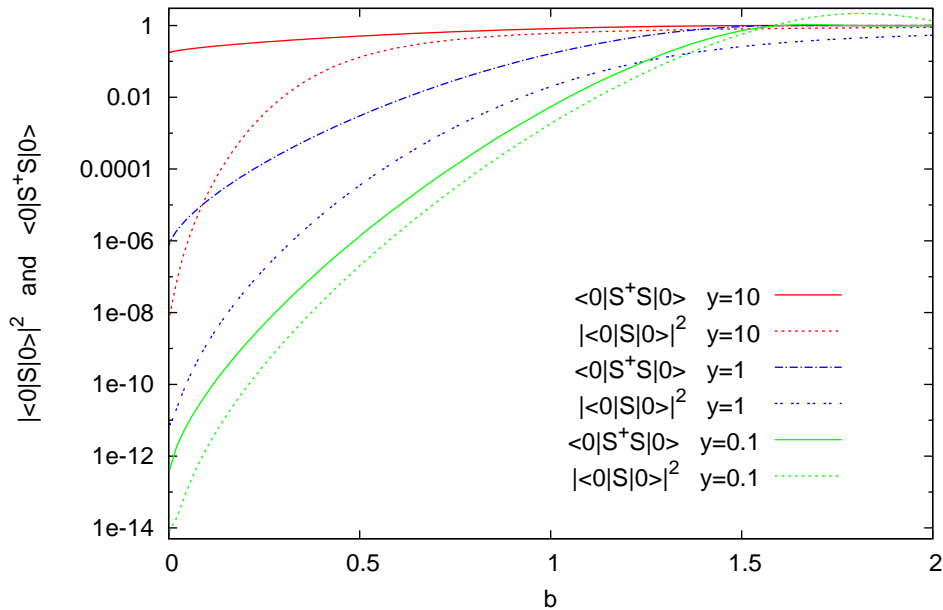}
  \caption{\it Comparison of the quantum v.e.v.~squared of the $S$-matrix
    (dashed lines) with the semiclassical v.e.v.~of the $S^\dagger S$ operator
    (solid lines) for $\alpha=5$ and various values of the absorption parameter
    y.}
  \label{f:sVSss}
\end{figure}

We have solved numerically the evolution equations~(\ref{eom12}) for
$(\rho_1,\rho_2)$, and we have obtained the unitarity action~(\ref{Au}) and the
semiclassical vacuum-expectation value of $S^\dagger S$~(\ref{0SS0sc}) for
different values of $y$.  In fig.~\ref{f:sVSss} we show our results for
$\alpha=5$, $y=0.1,\,1,\,10$, and compare them to the elastic quantum $S$-matrix
squared $|\bk{0|S|0}|^2 = \bk{0|\S^\dagger|0}\bk{0|\S|0}$ which gives the
vacuum-channel contribution to the unitarity sum~(\ref{unitSum}).  We shall
refer to the solutions for $|\bk{0|S|0}|^2$ as ``exclusive'' and to those of
$\bk{0|S^\dagger S|0}$ as ``inclusive'' over the inelastic states.

We note that, apart from the unphysical overshoot $|\bk{0|S|0}|>1$ of the
transition amplitude at small-$y$ and $b\gtrsim b_c$,%
\footnote{The small overshoot $|\bk{0|S|0}|^2 > 1$ at low $y = 0.1$ for $b
  \gtrsim 1.5$ is due to the oscillations of the quantum transition amplitude as
  seen in fig.~\ref{f:transAmpl}, compared to the semiclassical evaluation of
  $S^\dagger S$.}
the inequality $|\bk{0|S|0}|^2 \leq |\bk{0|S^\dagger S|0}|$ is always satisfied.
In the small-$y$ limit, inelastic effects are pretty small, in the sense that
$|\bk{0|S|0}|^2 \sim |\bk{0|S^\dagger S|0}|$. This reflects the fact that
$\rho_i(\tau,y)$ coincide with the vacuum solutions in the $y\to 0$
limit~(\ref{eomrho}).  Correspondingly, there is a sizeable unitarity violation
for $b < b_c \simeq 1.6$, inelastic effects providing corrections of relative
order $\ord{y}$.

\begin{figure}[ht!]
  \centering
  \includegraphics[angle=270,width=0.49\textwidth]{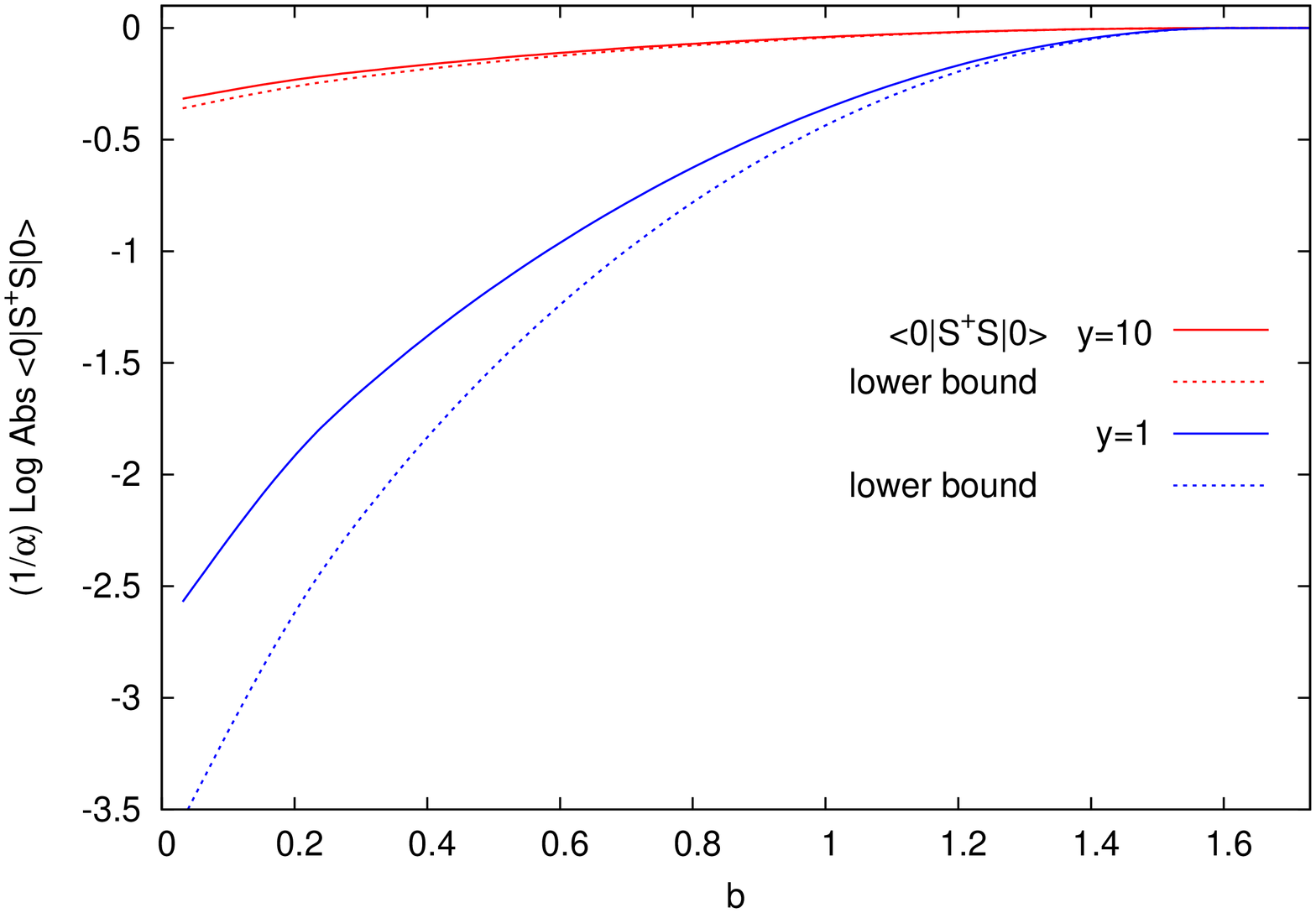}
  \hfill
  \includegraphics[angle=270,width=0.49\textwidth]{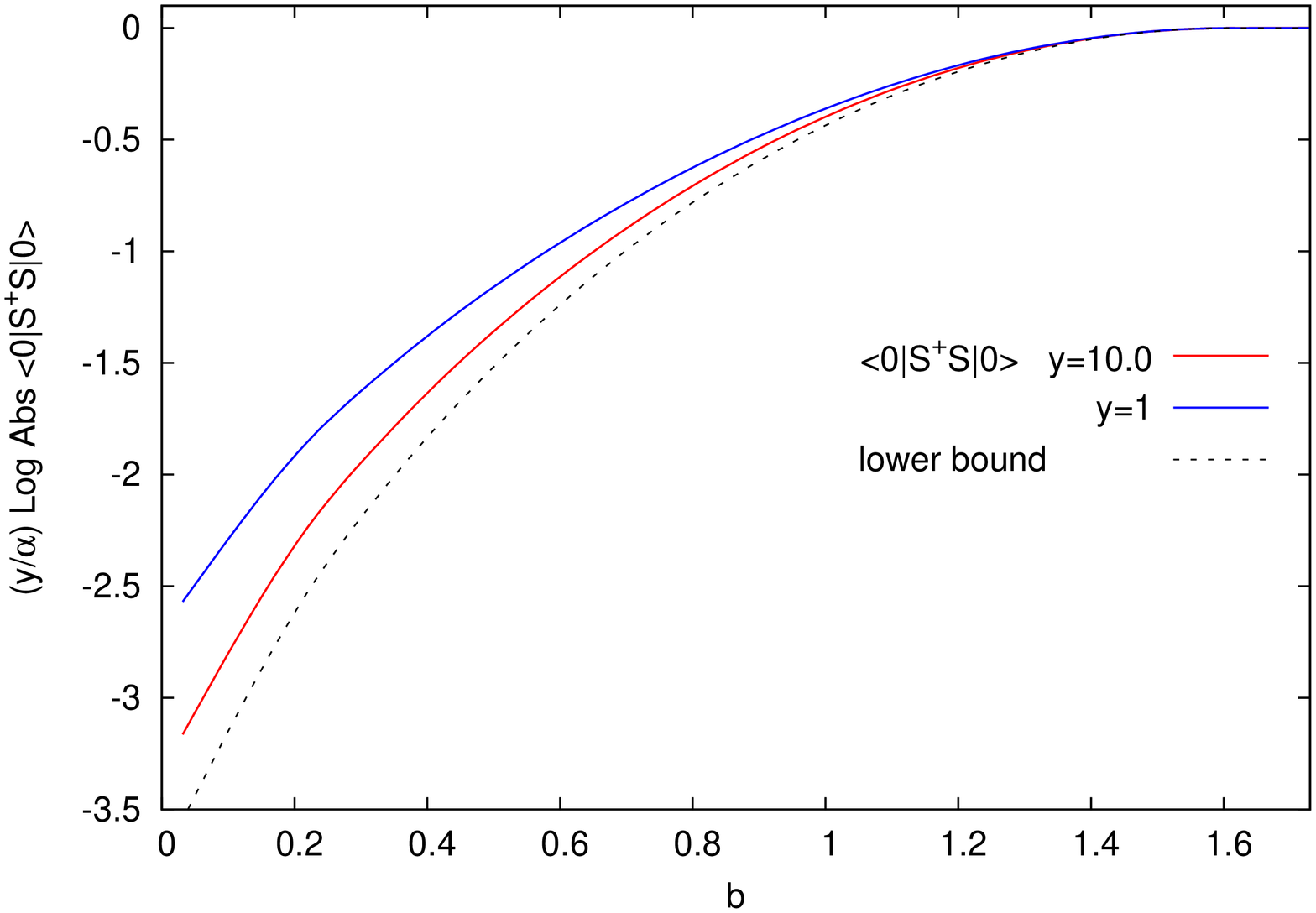}
  \caption{\it Comparison of the unitarity action (solid lines) with the
    unitarity bound estimates~(\ref{oRoRmin}) (dashed lines) for $\alpha=5$ and
    various values of the absorption parameter $y$. On the left we observe that
    the unitarity violation for $b < b_c \simeq 1.6$ vanishes for increasing
    values of $y$. On the right, we see that $y$ times the unitarity action
    tends to a finite limit, which is closely bounded from below by the
    estimate~(\ref{oRoRmin}).  }
  \label{f:unitarityBound}
\end{figure}

On the other hand, for large values of $y$, inelastic effects are very
important, and the $S$-matrix is approximately unitary. In this case, the
inclusive solutions are markedly different from the exclusive ones. The latter
scale as $\dot\rho(b^2,\tau,y)=\dot\rho\big(b^2(1-\ui y),\tau(1-\ui y),0)$ and
thus are peaked around $\tau\sim 1/y$, with $b^2 \sim b_c^2/y$, as roughly seen
in fig.~\ref{f:sVSss} so that the tunneling regime is displaced towards smaller
values of $b$. This implies in particular that the inelastic weight
[cfr.~eq.~(\ref{scomp})] $y\int h_2^2\; \dif\tau = \ord{1}$ thus showing the
importance of inelastic states yielding a finite (non-vanishing) contribution to
the unitarity sum~(\ref{unitSum}) in the large-$y$ limit.  The inclusive
solutions, instead, have $h_1\sim\ord{1}$ around $\tau=1$ and $h_2\sim\ord{1/y}$
everywhere, yielding a ``critical'' behaviour around $b\sim b_c$, as expected.
Since $h_2$ is small for large $y$ values, this implies that the on-shell
unitarity action scales as $\alpha/y$, yielding small unitarity violations in
this limit (figs.~\ref{f:sVSss},\ref{f:unitarityBound}).

The unitarity action is compared in fig.~\ref{f:unitarityBound} with the
unitarity sum~(\ref{udef},\ref{oRoRmin}) provided in the previous section. We
see that the latter is a good approximation to the unitarity action for large
$y$'s, thus providing some understanding of the coherent states dominating the
unitarity sum~(\ref{unitSum}), with the corresponding inelasticity $y$.

At this point, it becomes important to look at the $y\to\infty$ model, which is
unitary. Since $b_c^2(y)$ scales as $b_c^2(0)/y$, unitarity effects are mostly
seen in the small-$b$ region, as illustrated in fig.~\ref{f:sVSss}. We see that
for large $y$'s inelastic effects indeed fill the unitary defect. Note that
$|\bk{0|S|0}|\sim\esp{-\pi\alpha/2}$ in this case (instead of
$|\bk{0|S|0}|\sim\esp{-\pi\alpha}$ at $y=0$), thus showing that inelastic
effects compensate a finite unitarity defect around $b=0$, consistently with the
previous estimate of $y\int h_2^2\;\dif\tau$, providing the order of magnitude
of such effects.

\section{Discussion\label{s:disc}}

We have presented here a rather comprehensive study of a quantum extension of
the ACV gravitational $S$-matrix, both for the elastic matrix element (including
absorption) and for the inelastic ones. We have thus been able to provide an
analysis of the unitarity problem in the classical collapse region.

A striking outcome of the paper is that our $S$-matrix model satisfies
inelastic unitarity for all values of $b$ in the large-$y$ limit
$y\gg\alpha\gg1$. We all know how difficult it is to check unitarity, even in
well-known theories where no puzzling classical behaviour is present. Therefore,
this result is a quite non-trivial one and encourages us to further investigate
the large-$y$ model in detail in order to understand the features of the
inelastic production which is able to compensate the exponential tunneling
suppression in the small-$b$ region.

A key role, in recovering unitarity, is played by the quantum structure of our
$S$-matrix, which allows field fluctuations to build up a class of unitary
eigenstates, as explained in sec.~\ref{s:eesm}. Such states, characterized by
strong fields and small vacuum overlap at finite $y$'s, become actually dominant
in the $y \gg \alpha$ limit and turn out to saturate the unitarity sum.

On the other hand, the regime $y\gg\alpha=Gs/\hbar$ appears to be disfavoured
for $b<b_c$ on the basis of energy conservation and absorptive
corrections~\cite{CVprep}, because for $b\sim R$ emitted gravitons have a
somewhat hard transverse mass $\sim\hbar/R$, finally restricting $y$ to be at
most $\ord{\log\alpha}$ and actually $\ord{1}$ in the classical collapse
region.%
\footnote{Gravitons ($k$) are preferentially emitted in the large-angle region
  $\theta_k > \theta_q \simeq \hbar/bE$ ($q$ is the scattered particle), so that
  $Y \lesssim \log(Eb/\hbar)$ if the average graviton number
  $\langle n_g \rangle \leq 1$, or $Y\lesssim\log(Eb/\hbar\langle n_g \rangle)$
  if $\langle n_g \rangle>1$ (cfr.~ref.~\cite{CVprep}). By specializing to the
  collapse region $b\sim R$, we get the limitation above.}
This means that the unitarity defect that we find for finite $y$'s seems to be
the normal feature predicted by our model in the physically acceptable range of
$y$'s. An interesting point is that --- as we noted in sec.~\ref{s:uaf} --- it
is a defect and not an overflow.  A possible interpretation of that would be
that, in our quantum model, some information loss does show up in the classical
collapse region.

However, we do not really believe the unitarity defect of our model to be a
possible feature of a consistent quantum gravity theory. We rather think that
some of the approximations of string-gravity theory being used in building up
the model were inadequate.

Perhaps the weakest point of our model is the use of an uncorrelated coherent
state to represent inelastic production in the $S$-matrix for any given field
$h(\tau)$. From the original derivation~\cite{ACV07}, we know that
correlations are down by a power of $y$ (actually, a power of $\alpha y$) with
respect to uncorrelated emissions.  This hierarchy in $y$ could perhaps provide
a rationale for the need of a large-$y$ regime to recover unitarity.
Furthermore, the existence of correlations could provide a non-linear coherent
state, and thus a sort of ``condensation'' field which could change considerably
the analysis of saddle-points in the strong-field configurations and thus
provide a mechanism for recovering unitarity. We note that this non-linearity is
to some extent predictable from the diagrammatic approach of~\cite{ACV07}, based
on the multi-H diagrams of fig.~\ref{f:multiH}.

We further mention the fact that our quantization procedure keeps frozen the
longitudinal space-time structure of the shock-wave. This also is a weak point,
and correcting for it --- although much more difficult --- could provide again
further non-linearities in the reduced action and in the $S$-matrix coherent
state.

A different way of thinking is to believe that --- associated to the classically
collapsing states --- there are new quantum states, perhaps bound states, which
could contribute to the unitarity sum even if the explicit phase-space parameter
$y$ were set to zero. We have nothing in principle against this point of view,
we only find it difficult to implement it in a predictive way.

To sum up, our investigation of the quantum reduced-action model has led, in
part, to a conclusive answer, by exhibiting a unitary version of the model in
the (somewhat formal) large-$y$ limit. Future developments include the
understanding of the inelastic production of the unitary model which is
calculable within our approach. Furthermore, in order to possibly achieve
unitarity at finite values of $y$, we think we need improvements of the model
itself, probably in the direction of correlated emission, which looks important
at finite $y$'s in the classical collapse region.

\section*{Acknowledgements}

It is a pleasure to thank Daniele Amati and Gabriele Veneziano for a number of
discussions that helped us to find our way through the unitarity issue.

\appendix

\section{Eigenstates of the $\bs{S}$-matrix\label{a:esm}}

In this appendix we determine a set of eigenstates of the quantum $S$-matrix,
and argue that such set is complete in the Fock-space of gravitons.

The basic ideas are taken from the simpler analogue of a one-dimensional
harmonic oscillator with destruction and creation operators $a$ and $a^\dagger$
with the usual commutation relation $[a,a^\dagger]=1$. The bare-bone structure
of the $S$-matrix~(\ref{Scsh}) is in this case
\begin{equation}\label{S1d}
 S = \esp{\ui\Omega} \;, \qquad \Omega \equiv a+a^\dagger \;,
\end{equation}
where we note that $\Omega$ is proportional to the position operator.  An
eigenvector $\Ket{\omega}$ of $\Omega$ (and therefore of $S$) with eigenvalue
$\omega\in\R$ can be formally found by applying to any state $\ket{\psi}$ the
operator $\delta(\Omega-\omega)$:
\begin{equation}\label{formO}
  \Omega [\delta(\Omega-\omega) \ket{\psi}] = 
  \omega [\delta(\Omega-\omega) \ket{\psi}]
 \qquad \Rightarrow \qquad
  \Ket{\omega} = \delta(\Omega-\omega) \ket{\psi} \;.
\end{equation}
By using the vacuum state $\ket{\psi} = \ket{0}$ and the standard integral
representation of the Dirac delta, we find
\begin{equation}\label{ketOmega}
 \Ket{\omega} = \int_{-\infty}^{+\infty} \frac{\dif\zeta}{2\pi} \;
 \esp{-\ui\zeta\omega}
 \esp{\ui\zeta(a+a^\dagger)}\ket{0}
 \equiv \int\frac{\dif\zeta}{2\pi} \; \esp{-\ui\zeta\omega} \ket{\ui\zeta}
 \;.
\end{equation}
In words, the eigenstates of the position operator can be constructed as Fourier
transforms of coherent states
$\ket{\ui\zeta}\equiv\esp{\ui\zeta(a+a^\dagger)}\ket{0}$.  In particular,
$S\Ket{\omega} = \esp{\ui\omega}\Ket{\omega}$.

It is well known that the set of coherent states
$\ket{z}:z\in\C,\;a\ket{z}=z\ket{z}$ is (over) complete.  Actually, also the
subset of coherent states involved in eq.~(\ref{ketOmega}) with pure imaginary
eigenvalues $z=\ui\zeta$ is complete in the Hilbert space $H$.  In fact, the map
$z\mapsto\esp{z\,a^\dagger}\ket{0} = \esp{|z|^2/2}\ket{z}$, $\C\to H$ is
holomorphic, and thus any coherent state $\ket{z_0}$ can be represented as a
superposition of ``pure imaginary'' coherent states according to the Cauchy
integral
\begin{equation}\label{imagCS}
 \esp{z_0 a^\dagger}\ket{0} = -\sign(\Re(z_0)) \;
 \lim_{\epsilon\to0} \int\frac{\dif z}{2\pi\ui} \;
 \frac{\esp{\epsilon z}}{z-z_0} \esp{z a^\dagger}\ket{0}
\end{equation}
where $z=\ui\zeta$ runs along the imaginary axis and the sign of $\epsilon$ is
opposite to the sign of $\Re(z_0)$ in such a way that the integration path can
be closed around $z_0$.

Coming back to the infinite-dimensional Hilbert space of gravitons with the
destruction and creation operators $A(\xt)$ and $A^\dagger(\xt)$ in
eq.~(\ref{Omega}), we observe that the $S$-matrix~(\ref{inSpi}) involves an
azimuthally invariant integration of $A(\xt)+A^\dagger(\xt)$. It is therefore
convenient to introduce the canonically normalized operators
\begin{equation}\label{azimA}
 a(\tau=\xt^2) \equiv \int_0^{2\pi} \frac{\dif\phi_{\xt}}{2\sqrt{\pi}} \;
 \frac{A(\xt)}{\sqrt{Y}} \qquad\Rightarrow\qquad
 [a(\tau),a^\dagger(\tau')] = \delta(\tau-\tau') \;,
\end{equation}
whose eigenstates are coherent states depending on a functional parameter
$\eta(\tau)\in\C$:
\begin{equation}\label{infCS}
 \ket{\eta(\tau)} \equiv \esp{(\eta,a^\dagger)-(\eta^*,a)}\ket{0}
 = \esp{-\frac12(\eta,\eta)} \esp{(\eta,a^\dagger)}\ket{0} \;, \qquad
 a(\tau) \ket{\eta(\tau')} = \eta(\tau)\ket{\eta(\tau')}
\end{equation}
with the scalar product notation
$(\eta,\zeta)\equiv\int_0^\infty\eta(\tau)\zeta(\tau)\;\dif\tau$.
We argue, by analogy with the one-dimensional case, that the set of coherent
states with pure imaginary functional parameter $\eta(\tau) = \ui\zeta(\tau)$,
$\zeta(\tau)\in\R$, is complete in the Fock space of gravitons.

With the notations above, the $S$-matrix~(\ref{inSpi}) can be written in the
compact form
\begin{equation}\label{Scompact}
 S = \int  \Dif{\rho(\tau)} \;
 \esp{-\ui \int L(\rho)\;\dif\tau +\ui(\delta_\rho,a+a^\dagger)} \;, \qquad
 \delta_\rho \equiv \sqrt{2\alpha y}(1-\dot\rho) \;.
\end{equation}
By using the Backer-Campbell-Hausdorff relations
\begin{equation}\label{bch}
 \esp{(\eta,a)+(\tilde\eta,a^\dagger)} = \esp{\frac12(\eta,\tilde\eta)}
 \esp{(\tilde\eta,a^\dagger)} \esp{(\eta,a)} \;, \qquad
 \esp{(\eta,a)} \esp{(\tilde\eta,a^\dagger)} = \esp{(\eta,\tilde\eta)}
 \esp{(\tilde\eta,a^\dagger)} \esp{(\eta,a)} \;,
\end{equation}
for casting operators in normal ordering, we can easily derive the action of the
$S$-matrix on the coherent states:
\begin{align}
 S\ket{\ui\zeta(\tau)} &= \int\Dif{\rho(\tau)} \; \esp{-\ui\int L(\rho)}
 \esp{-\frac12(\delta_\rho,\delta_\rho)}
 \esp{\ui(\delta_\rho,a^\dagger)} \esp{\ui(\delta_\rho,a)}
 \esp{-\frac12(\zeta,\zeta)} \esp{\ui(\zeta,a^\dagger)} \ket{0}
 \nonumber \\
 &= \int\Dif{\rho(\tau)} \; \esp{-\ui\int L(\rho)}
 \esp{-\frac12(\zeta+\delta_\rho,\zeta+\delta_\rho)}
 \esp{\ui(\zeta+\delta_\rho,a^\dagger)} \ket{0}
 = \int\Dif{\rho(\tau)} \; \esp{-\ui\int L(\rho)} \ket{\ui(\zeta+\delta_\rho)} \;.
 \label{Szeta}
\end{align}
In practice, for each path $\rho(\tau)$, the coherent state parameter
$\zeta(\tau)$ is shifted by an amount $\delta_\rho(\tau)$.%
\footnote{This motivates the notation $\delta_\rho$ in the
  definition~(\ref{Scompact}).}

In order to look for eigenstates of the $S$-matrix, we introduce the functional
Fourier transform of coherent states
\begin{equation}\label{funcFT}
 \Ket{\omega(\tau)} \equiv N \int\Dif{\zeta(\tau)} \; \esp{-\ui(\zeta,\omega)}
 \ket{\ui\zeta(\tau)} \;,
\end{equation}
where $N$ is a normalization factor which can be determined by computing
\begin{align}
 \langle\{\omega'(\tau)\}|\{\omega(\tau)\}\rangle &=
 N'{}^* N \int\Dif{\zeta'(\tau)}\Dif{\zeta(\tau)}\; \esp{-\ui(\omega',\zeta')
 +\ui(\omega,\zeta) - \frac12(\zeta'-\zeta,\zeta'-\zeta)} \nonumber \\
 &= N'{}^* N \esp{-\frac12(\omega',\omega')} \int\Dif{\zeta(\tau)} \;
 \esp{\ui(\zeta,\omega-\omega')}
 = |N|^2 \esp{-\frac12(\omega,\omega)} \delta(\{\omega-\omega'\})
\end{align}
thus requiring $N=\esp{\frac14(\omega,\omega)}$ for $\Ket{\omega(\tau)}$ to be a
complete and orthonormal set.

This set diagonalizes the $S$-matrix operator. In fact, by using
eqs.~(\ref{Szeta},\ref{funcFT}) we find
\begin{align}
 S\Ket{\omega(\tau)} &= N\int\Dif{\rho(\tau)}\Dif{\zeta(\tau)} \;
 \esp{-\ui\int L(\rho)} \esp{-\ui(\zeta,\omega)} \ket{\ui(\zeta+\delta_\rho)}
 \nonumber \\
 &= \int\Dif{\rho(\tau)} \; \esp{-\ui\int L(\rho)+\ui(\delta_\rho,\omega)}
 \; \Ket{\omega(\tau)}
\end{align}
where we have decoupled the two integrations by shifting
$\zeta\to\zeta'=\zeta+\delta_\rho$.  The eigenvalue of the $S$-matrix relative
to the eigenstate $\Ket{\omega(\tau)}$ is expressed by a path-integral in $\rho$
\begin{align}
 \text{eigenv}_\omega(S) &\equiv \esp{\ui\A[\omega]} = \int\Dif{\rho(\tau)} \;
  \esp{-\ui\int L(\rho)+\ui(\delta_\rho,\omega)} \nonumber
\end{align}
which can be estimated in the semiclassical approximation by finding the path
$\rho_\omega(\tau)$ around which the ``action'' $\A[\omega]$ is stationary, as
explained in sec.~(\ref{s:eesm}).

\section{The unitarity action\label{a:ua}}

In this section we compute the unitarity action~(\ref{d:Au}) corresponding to
the stationary/classical trajectory determined, for $b<b_c$, by the equation of
motion~(\ref{eom12}) and boundary conditions~(\ref{bc}).
In terms of the real components $(\rho_1,\rho_2)$ defined by
$\rho\equiv\rho_1+\ui\rho_2=\tilde{\rho}^*$, the unitarity action reads
\begin{subequations}\label{ua12}
  \begin{align}
    \A_u &= -2\ui\alpha \int_0^\infty (2\dot\rho_1 \dot\rho_2 - 2\dot\rho_2 +
    2y\dot\rho_2^2 - V_u) \; \dif\tau
    \label{L12} \\
    V_u(\rho_1,\rho_2;\tau) &\equiv \Theta(\tau-b^2) \Im\frac1{\rho}
    = \Theta(\tau-b^2) \frac{\rho_1-\ui\rho_2}{\rho_1^2+\rho_2^2} \;.
    \label{Vu}
  \end{align}
\end{subequations}

In the interval $0<\tau<b^2$, the potential $V_u$ vanishes. Therefore, the
equation of motions $\ddot\rho_1 = \ddot\rho_2 = 0$ determine a free evolution
for the $\rho$ field, whose solution is $\rho_k(\tau) = t_k\tau,\;(k=1,2)$,
where the $t_k\equiv\dot\rho_k(0)$ are free parameters (eventually constrained
by the boundary conditions at $\tau=\infty$), having taken into account the
initial condition $\rho_k(0) = 0$. The corresponding contribution to the action
amounts to
\begin{equation}\label{Au1}
 \A_u|_{\tau<b^2} = -4\ui\alpha b^2 [ t_1t_2
 -t_2 + y t_2^2 ] \;.
\end{equation}

In the interval $\tau>b^2$ the evolution is nontrivial, and we need some
relations among the $\rho_k$'s and their $\tau$-derivatives. Since the
``unitarity lagrangian'' in eq.~(\ref{ua12}) is time-independent for $\tau>b^2$,
the corresponding hamiltonian
\begin{equation}\label{Hu}
 H_u = 2\ui [ 2\dot\rho_1\dot\rho_2+2y\dot\rho_2^2 + V_u] = 0
\end{equation}
is a constant of motion, and evaluates to zero because of the boundary condition
$\dot\rho(\infty)=1$ that implies
$\dot\rho_1(\infty)=1,\,\dot\rho_2(\infty)=0,\,V_u(\infty)=0$.
Another useful relation is obtained by multiplying the first equation
of~(\ref{eom12}) by $\rho_2$ and the second one by $\rho_1$, yielding
\begin{equation}\label{rel2a}
 2\ddot\rho_1\rho_2 + 2\rho_1\ddot\rho_2 + 4y\rho_1\ddot\rho_2
 = \Re\frac1{\rho^2}\Im\rho + \Im\frac1{\rho^2}\Re\rho
 = \Im\frac1{\rho} = V_u \;.
\end{equation}
In turn, by using the identities
$(\rho_1\rho_2)\,\ddot{}\,=\ddot\rho_1\rho_2+\rho_1\ddot\rho_2+2\dot\rho_1\dot\rho_2$,
$(\rho_2^2)\,\ddot{}\,=2\rho_2\ddot\rho_2+2\dot\rho_2^2$ and the integral of
motion~(\ref{Hu}), we obtain
\begin{equation}\label{rel2}
 2(\rho_1\rho_2+y\rho_2^2)\,\ddot{}\, + V_u = 0 \;.
\end{equation}
The action for $\tau>b^2$ can now be computed:
\begin{align}
 \A_u|_{\tau>b^2} &\stackrel{(\ref{Hu})}{=} -2\ui\alpha
 \int_{b^2}^\infty (-2\dot\rho_2-2V_u)\;\dif\tau \nonumber\\
 &\stackrel{(\ref{rel2})}{=}4\ui\alpha\int_{b^2}^\infty
 [\dot\rho_2-2(\rho_1\rho_2+y\rho_2^2)\,\ddot{}\,]\;\dif\tau \nonumber\\
 &\;\,= -4\ui\alpha[\rho_2(b^2)-\rho_2(\infty)+2(\rho_1\rho_2+y\rho_2^2)
 \,\dot{}\,(\infty)-2(\rho_1\rho_2+y\rho_2^2)\,\dot{}\,(b^2)] \;.
 \label{Au2}
\end{align}
The values of $\rho_k(b^2)$ and of its derivatives are matched with those of the
free solution for $\tau\leq b^2$. At $\tau\to\infty$ we have
$\rho_1=\ord{\tau}$, $\rho_2=\ord{1}$,
$\ddot\rho_2\sim-2\rho_2/\rho_1^3=\ord{\tau^{-3}}$,
$\dot\rho_2=\ord{\tau^{-2}}$, hence
$(\rho_1\rho_2+y\rho_2^2)\,\dot{}\, \to \rho_2(\infty)$.

By summing the results~(\ref{Au1},\ref{Au2}) we obtain
\begin{equation}\label{AuTot}
 \A_u = -4\ui\alpha[\rho_2(\infty)-3 b^2 t_2(t_1+y t_2)]
 = -4\ui\alpha\left[\rho_2(\infty) -\frac32\frac{t_2}{t_1^2+t_2^2}\right] \;,
\end{equation}
where in the last equality we exploited the relation
\begin{equation}\label{imbq}
 2 t_2 ( t_1 + y t_2) = -V_u(b^2_+) = \frac{t_2}{b^2(t_1^2+t_2^2)} \;.
\end{equation}
obtained from the $\tau\to b^2_+$ limit of the integral of motion~(\ref{Hu}).

\subsection{$\boldsymbol{y\to\infty}$ limit\label{s:yinf}}

The boundary problem defined in eqs.~(\ref{eom12},\ref{bc}) admits a well
defined limit for $y\to\infty$. In fact, by setting
$R_2(\tau)\equiv y\rho_2(\tau)$, we obtain
\begin{align}
  &\begin{cases}
    2\ddot\rho_1 + 4 \ddot R_2 &=
    \displaystyle{\frac{\rho_1^2-R_2^2/y^2}{(\rho_1^2+R_2^2/y^2)^2}
      \Theta(\tau-b^2)\quad \to \frac1{\rho_1^2}\Theta(\tau-b^2)} \\[2ex]
    2\ddot R_2 &= \displaystyle{-\frac{2\rho_1 R_2}{(\rho_1^2+R_2^2/y^2)^2}
      \Theta(\tau-b^2) \;\to -\frac{2 R_2}{\rho_1^3}\Theta(\tau-b^2)}
  \end{cases}
 \label{Req} \\
 &\rho_1(0) = 0 \;, \quad R_2(0) = 0 \;, \quad
 \dot\rho_1(\infty) = 1 \;, \quad \dot{R}_2(\infty) = 0 \;.
 \label{Rbc}
\end{align}
The above system has a finite solution for the pair of functions $(\rho_1,R_2)$
in the $y\to\infty$ limit. We deduce that, at large $y$, the real part $\rho_1$
of $\rho$ tends to a finite limit, whereas the imaginary part $\rho_2$ of $\rho$
uniformly scales as $\frac1{y} R_2^{[y=\infty]}$. Therefore, the
quantities $\rho_2(\infty)$, $t_2$ and $\A_u$ linearly vanishes with $1/y$.  The
fact that $\lim_{y\to\infty}\A_u=0$ suggests the unitarity of the model at
$y=\infty$.


\end{document}